%% file: qg-paper.tex
\documentclass[submission, Phys]{SciPost}

\pdfoutput=1
\input{incl_settings.tex}
\input{incl_shortcuts.tex}
\graphicspath{{./figures/}}

\begin{document}

\begin{center} 
    {\Large\textbf{Performance versus Resilience in Modern Quark--Gluon Tagging}}
\end{center}

\begin{center}
    Anja~Butter\textsuperscript{1,2},
    Barry~M.~Dillon\textsuperscript{1},
    Tilman~Plehn\textsuperscript{1} and
    Lorenz~Vogel\textsuperscript{1}
\end{center}

\begin{center}
    {\bf 1} Institut f\"{u}r Theoretische Physik, Universit\"{a}t Heidelberg, Germany\\
    {\bf 2} LPNHE, Sorbonne Universit\'{e}, Universit\'{e} Paris Cit\'{e}, CNRS/IN2P3, Paris, France
\end{center}

\begin{center}
    \today
\end{center}
 
\section*{Abstract}
{\bf Discriminating quark-like from gluon-like jets is, in many ways,
  a key challenge for many LHC analyses. First, we use a known
  difference in \pythia and \herwig simulations to show how
  decorrelated taggers would break down when the most distinctive
  feature is aligned with theory uncertainties. We propose
  conditional training on interpolated samples, combined with a
  controlled Bayesian network, as a more resilient framework. The
  interpolation parameter can be used to optimize the training
  evaluated on a calibration dataset, and to test the stability of
  this optimization. The interpolated training might also be
  useful to track generalization errors when training  networks
  on simulations.}

\vspace{10pt}
\noindent\rule{\textwidth}{1pt}
\tableofcontents\thispagestyle{fancy}
\noindent\rule{\textwidth}{1pt}
\vspace{10pt}

\clearpage
%%%%%%%%%%%%%%%%%%%%%%%%%%%%%%%%%%%%%%%%%%%%%%%%%%%%%%%%%%%%%%%%%%%%%%%%%%%%%%
\section{Introduction}
\label{sec:intro}

Jets are the main analysis objects at the LHC, and the success of the
LHC program is, to a large degree, being driven by an improved
understanding of jets experimentally and theoretically. In practice,
the main task in jet physics is to predict their features precisely,
and to use them to tag the parton initiating the jet. The improved
understanding of subjet physics at the LHC has allowed us to skip
high-level observables and instead analyze jets using low-level
detector output employing modern machine
learning~\cite{Plehn:2022ftl}. ML-tools can incorporate all available
information in a jet and significantly improve the performance of
classic, multivariate jet taggers. While, in the interest of
optimality, taggers should be trained on data, ATLAS and CMS follow a
more conservative approach and train taggers on simulations. This
attitude reflects a flattering trust in theoretical simulations, but
it also creates new sources of uncertainties.

In this paper we add uncertainty-aware features to the CMS ParticleNet
tagger~\cite{Qu:2019gqs}, using a Bayesian classification network
setup~\cite{Bollweg:2019skg}, and propose an interpolated training method
using conditional networks. This combination allows us to
capture different sources of uncertainties~\cite{Butter:2021csz} while
protecting the performance of the tagger. The Bayesian setup 
would raise a flag when the training datasets are too inconsistent 
to be combined. The conflict between optimal
performance and uncertainty control is the key weakness of adversarial
training approaches and makes it preferable to instead use nuisance
parameters to describe systematics~\cite{Ghosh:2021roe}. For theory 
uncertainties, adversarial approaches are not even likely to cover the
full uncertainty range~\cite{Ghosh:2021hrh}.

ML-methods for jet tagging~\cite{deOliveira:2015xxd} can be applied to
the whole range of top jets, Higgs jets, $W/Z$-jets, $\tau$-jets,
bottom or charm jets, all the way to quark \vs gluon jets. From a
theoretical and an experimental perspective it is easiest to tag
partons of which only the decay products hadronize. For instance top
taggers then look for distinctive features like the jet mass or the
multiplicity of subjet constituents~\cite{Kasieczka:2017nvn,
  Butter:2017cot, Macaluso:2018tck}. Being a well-defined problem, top
tagging has played a key role in developing and establishing a wide
range of network architectures~\cite{Kasieczka:2019dbj}, including
uncertainty-aware extensions.

In contrast, quark \vs gluon tagging is not actually defined
theoretically, beyond leading-order in QCD and including parton
splittings. Still, because of its great analysis potential,
quark--gluon tagging has a long history~\cite{Field:1977fa,
  Nilles:1980ys, Jones:1988ay, Fodor:1989ir, Jones:1990rz,
  Lonnblad:1990qp, CLEO:1992fdq, OPAL:1996irm}, 
including early applications at the LHC~\cite{ATLAS:2014vax,
  ATLAS:2017nma, ATLAS:2017dfg, CMS:2013kfa, CMS:2018sah,
  Andrews:2019faz}. In spite of the serious theoretical
challenges~\cite{Sterman:1977wj, Ellis:1993tq,
  Seymour:1997kj,Catani:1997xc, Banfi:2006hf, Salam:2010nqg,
  Gras:2017jty, Komiske:2018vkc, Larkoski:2019nwj, Romero:2021qlf},
efficient ML-approaches have been devised to separate ``quark jets''
from ``gluon jets''~\cite{Gallicchio:2012ez,Komiske:2016rsd,
  Cheng:2017rdo,Luo:2017ncs,Fraser:2018ieu}. They include study of
hadronization and detector effects~\cite{Kasieczka:2018lwf} and modern
network architecture like transformers~\cite{Mikuni:2021pou},
Lorentz-equivariant networks~\cite{Gong:2022lye}, and normalizing
flows~\cite{Dolan:2022ikg}. One way to overcome the fundamental
problem of defining quark and gluon jets is to instead use
well-defined hypotheses in terms of LHC signatures, for instance
mostly quarks in LHC signals like weak boson fusion \vs gluons in QCD
backgrounds~\cite{Gallicchio:2011xc, Bright-Thonney:2018mxq,
  Biekotter:2017gyu, Sakaki:2018opq, Chiang:2022lsn}. An alternative
method is to train a classifier without labels, just on samples with
an enhanced partonic quark or gluon fraction~\cite{Metodiev:2017vrx}.

In this paper we first look at a known issue, namely the differences
between \herwig and \pythia jets and the effect of these differences
on ML-taggers introduced in \cref{sec:specs_qg}. To control the
cutting-edge ParticleNet tagger and understand its output better, we
present its Bayesian variant in \cref{sec:specs_net}. It allows us to
understand the problem of quark--gluon taggers trained on \herwig
and \pythia and makes it obvious that a naive resilience improvement
through decorrelation will massively hurt the performance of the
tagger, as discussed in \cref{sec:where}. In \cref{sec:resilient} we
target this problem through a new, interpolated training of the
conditional ParticleNet tagger on two distinct samples. We realize
this interpolation with the same ParticleNet classifier. After
discussing this method in detail, we extend it to a fresh look at a
more interpretable, continuous calibration of jet taggers.

%%%%%%%%%%%%%%%%%%%%%%%%%%%%%%%%%%%%%%%%%%%%%%%%%%%%%%%%%%%%%%%%%%%%%%%%%%%%%%
\section{Dataset and classification network}
\label{sec:specs}

One of the most exciting goals of subjet tagging is the discrimination
of quarks versus gluons. The precise task is not well-defined beyond
leading order in QCD, but it approximates important questions like how
to identify electroweak decay jets or how to separate weak boson
fusion from QCD backgrounds. In both cases, the signals are
quark-enriched, while most QCD jets at the LHC come from gluon
emission. Another aspect which makes quark--gluon tagging especially
interesting is that there exists a study which raises questions about
the behavior of ML-taggers in this application.

%%%%%%%%%%%%%%%%%%%%%%%%%%%%%%%%%%%%%%%%%%%%%%%%%%%%%%%%%%%%%%%%%%%%%%%%%%%%%%
\subsection{Quark--gluon datasets}
\label{sec:specs_qg}

The starting point of our study are two datasets of simulated quark
and gluon jets~\cite{Komiske:2018cqr, Komiske:2019qgp,
  Pathak:2019qgh}, each with $2$M jets, one generated with \pythia and
one generated with \herwig. The two samples are generated using the
partonic processes
\begin{equation}
	q\bar{q}\to Z(\to\nu\bar{\nu})+g
	\qquad\mand\qquad 
	qg\to Z(\to\nu\bar{\nu})+(uds)
	\label{eq:sim_proc}
\end{equation}
at the $\SI{14}{\tera\electronvolt}$ LHC, simulated with
\pythia~8.226~\cite{Sjostrand:2006za, Sjostrand:2014zea} and with
\herwig~7.1.4~\cite{Bellm:2017bvx}. Both setups use default tunes and
shower parameters. Hadronization and multi-parton interactions (MPI)
are turned on, and we do not consider jets with charm or bottom quark
content. The jets are defined through anti-$\kt$
algorithm~\cite{Cacciari:2008gp, Cacciari:2005hq} in \fastjet
3.3.0~\cite{Cacciari:2011ma} with a radius of $R=0.4$. No detector
simulation is included, which cuts into the realism of the analysis,
but allows us to extract the underlying question and issues and solve
them before adding detector simulations to the problem. For each
event the dataset keeps the leading jet, provided
\begin{equation}
	\ptjet = \range{500}{550}\,\si{\giga\electronvolt}
	\qquad\mand\qquad
	\abs{\etajet} < 1.7 \eqperiod
\end{equation}
If we assume that all light-flavor jet constituents are approximately
massless, each jet $x_{i}$ is defined by
\begin{equation}
	x_{i} = \bigbraces{(\pt, \eta, \phi)_{k}}
	\qquad\mwith\qquad 
	k = \crange{1}{\nc} \eqperiod
\end{equation}
For our analysis we allow for up to $\nc=100$ constituents per jet. The
jets are zero-padded with constituents, and all constituents have
azimuthal angles $\phi$ within $\pi$ of the jet.

We refer to the final-state jets from the two partonic processes in
\cref{eq:sim_proc} as quark and gluon jets, even through it is
clear that this statement is scale dependent and only defined at
leading order in perturbation theory. A more appropriate way of
referring to these jets would be in the sense of semi-supervised
learning and quark-enhanced \vs gluon-enhanced samples. A standard way
of realizing this setup would be jets reconstructed as coming from a
two-body $Z$-decay \vs jets produced in association with a Higgs
boson.

We supplement the \pythia and \herwig datasets with a third simulation
of the two processes in \cref{eq:sim_proc} using
\sherpa~2.2.10~\cite{Sherpa:2019gpd}, again with the default tune and
shower parameters. Using \textsc{pyhepmc}~\cite{Rodrigues:2020syo,
  Dembinski2022}, a \python wrapper for the
\textsc{HepMC2}~\cite{Buckley:2019xhk} library, we select the
constituent coordinates of those final-state particles not labelled as
neutrinos. The \sherpa jets are defined through the
\textsc{PyJet}~\cite{Rodrigues:2020syo, Dawe2022} interface to
\fastjet. From a physics perspective, the \sherpa jet resembles the
\herwig jets through the common use of cluster fragmentation, but we
will see that the numerical results differ.

Each of our three jet datasets consists of $20$ files with $100$k jets
each, equally split between quark and gluon jets. For each generator
we divide the dataset into training/validation/test subsets with
$200$k/$50$k/$50$k jets for quarks and gluons, each, unless mentioned
otherwise.

For state-of-the-art jet tagging we need to include particle
identification (PID) information. Our \pythia and \herwig datasets
include two forms of PIDs~\cite{Komiske:2018cqr}, (i) the full
particle-ID information from \pythia or \herwig, and (ii)
experimentally realistic particle IDs. We follow the ParticleNet
approach~\cite{Qu:2019gqs}, using the five particle types electron,
muon, charged hadron, neutral hadron, and photon, plus the electric
charge as input to the network. The standard encoding by the Particle
Data Group in terms of large and irregular integer values is not an
ideal ML-input. Instead, we use a one-hot encoding of our
experimentally realistic PIDs.

%-----------------------------------------------------------
\begin{figure}[t]
    \centering
	\begin{subfigure}{0.50\linewidth}
		\centering
		\includegraphics[width=0.95\linewidth]{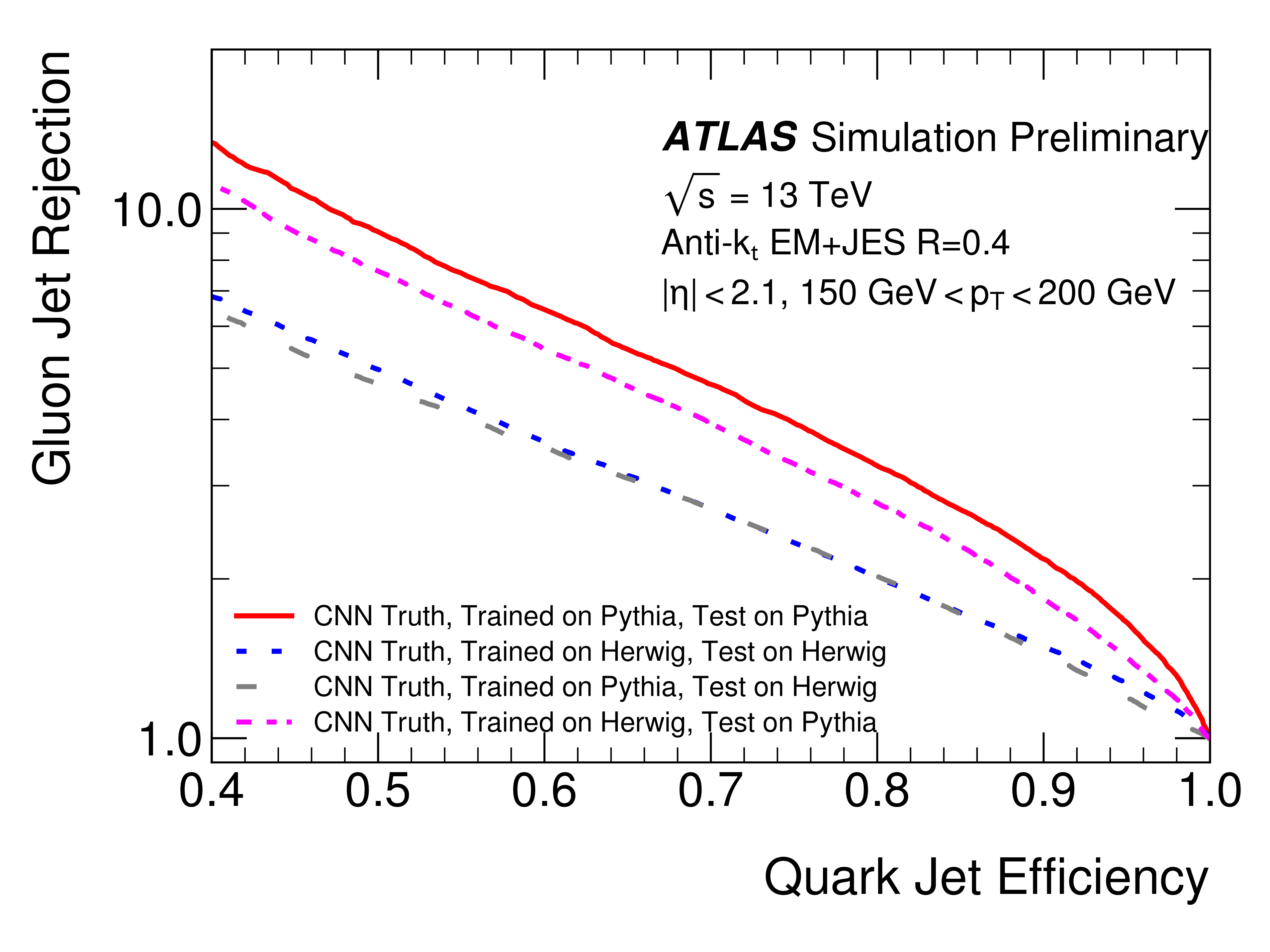}
		%\caption{}\label{fig:}
	\end{subfigure}\hfill
	\begin{subfigure}{0.50\linewidth}
		\centering
		\includegraphics[width=0.95\linewidth]{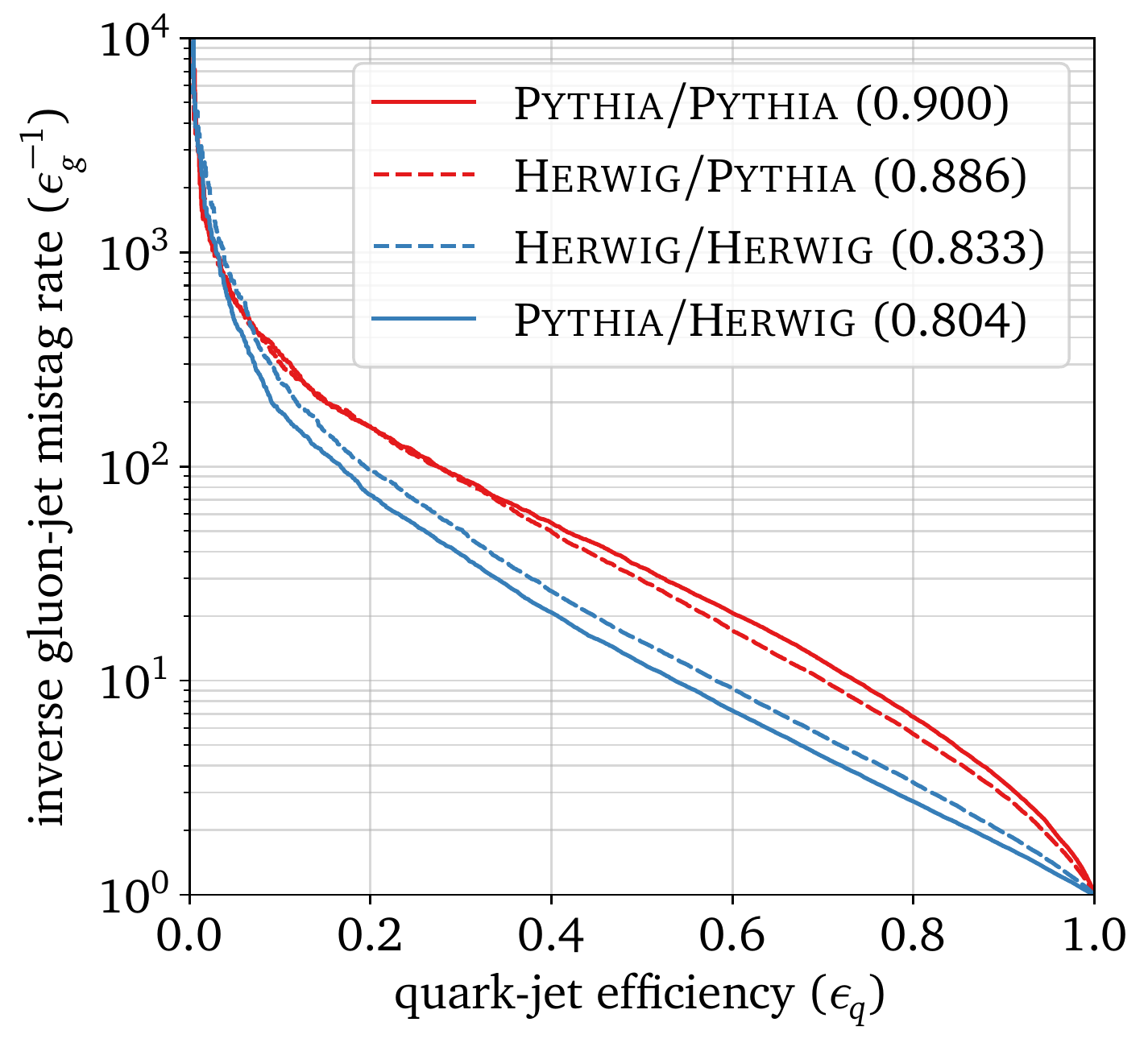}
		%\caption{}\label{fig:}
	\end{subfigure}
     \caption{\textit{Left:} preliminary result of the ATLAS study on quark--gluon tagging, raising questions on the best way to train such a tagger. Figure from Ref.~\cite{ATLAS:2017dfg}. The same pattern has been observed in Fig.~8 of Ref.~\cite{Komiske:2016rsd}. \textit{Right:} our results on the same task, using a new, Bayesian version of ParticleNet-Lite~\cite{Qu:2019gqs}, trained on 400k \pythia and
     \herwig jets each, with the parameters given in \cref{tab:bpn-arch}.}
     \label{fig:qg-weirdness}
\end{figure}
%-----------------------------------------------------------

%%%%%%%%%%%%%%%%%%%%%%%%%%%%%%%%%%%%%%%%%%%%%%%%%%%%%%%%%%%%%%%%%%%%%%%%%%%%%%
\subsubsection*{High-level observables}

There exist standard kinematic observables for subjet physics,
specifically quark--gluon discrimination~\cite{Biekotter:2017gyu},
for instance the multiplicity of constituents or particle flow objects
($\npf$), the radiation distribution or girth
($\wpf$)~\cite{Gallicchio:2010dq,Gallicchio:2011xq}, the width of the
$\pt$-distribution of the constituents ($\ptd$)~\cite{CMS:2013kfa}, or
the weighted angular correlator ($C_{0.2}$)~\cite{Larkoski:2013eya}.
They are defined in terms of the jet constituents as
\begin{equation}
	\begin{aligned}
		\npf &= \sum_{i} 1 &\qqqquad
		\wpf &= \frac{\sum_{i}\pti\Delta R_{i,\mathrm{jet}}}{\sum_{i}\pti} \\
		\ptd &= \frac{\sqrt{\sum_{i}\pti^{2}}}{\sum_{i}\pti} 
		& C_{0.2} &= \frac{\sum_{i, j} \pti\ptj(\Delta R_{ij})^{0.2}}{
			\bigl(\sum_{i}\pti\bigr)^{2}} \eqperiod
	\end{aligned}
	\label{eq:high_level}
\end{equation}
Distinguishing quark jets from gluon jets exploits two features
encoded in these observables~\cite{Kasieczka:2018lwf,Frye:2017yrw}.
First, the QCD color factors for quarks are smaller than for gluons,
which means radiating a gluon off a hard gluon versus off a hard quark
comes with the ratio $C_{A}/C_{F}=9/4$. This leads for a higher
multiplicity and broader girth for hard gluons. Second, the quark and
gluon splitting functions differ in the soft limit. The harder
fragmentation for quarks leads to quark jet constituents carrying a
larger average fraction of the jet energy, tracked by $\ptd$.

%-----------------------------------------------------------
\begin{figure}[t]
	\centering
	\begin{subfigure}{0.50\linewidth}
		\centering
		\includegraphics[width=0.94\linewidth]{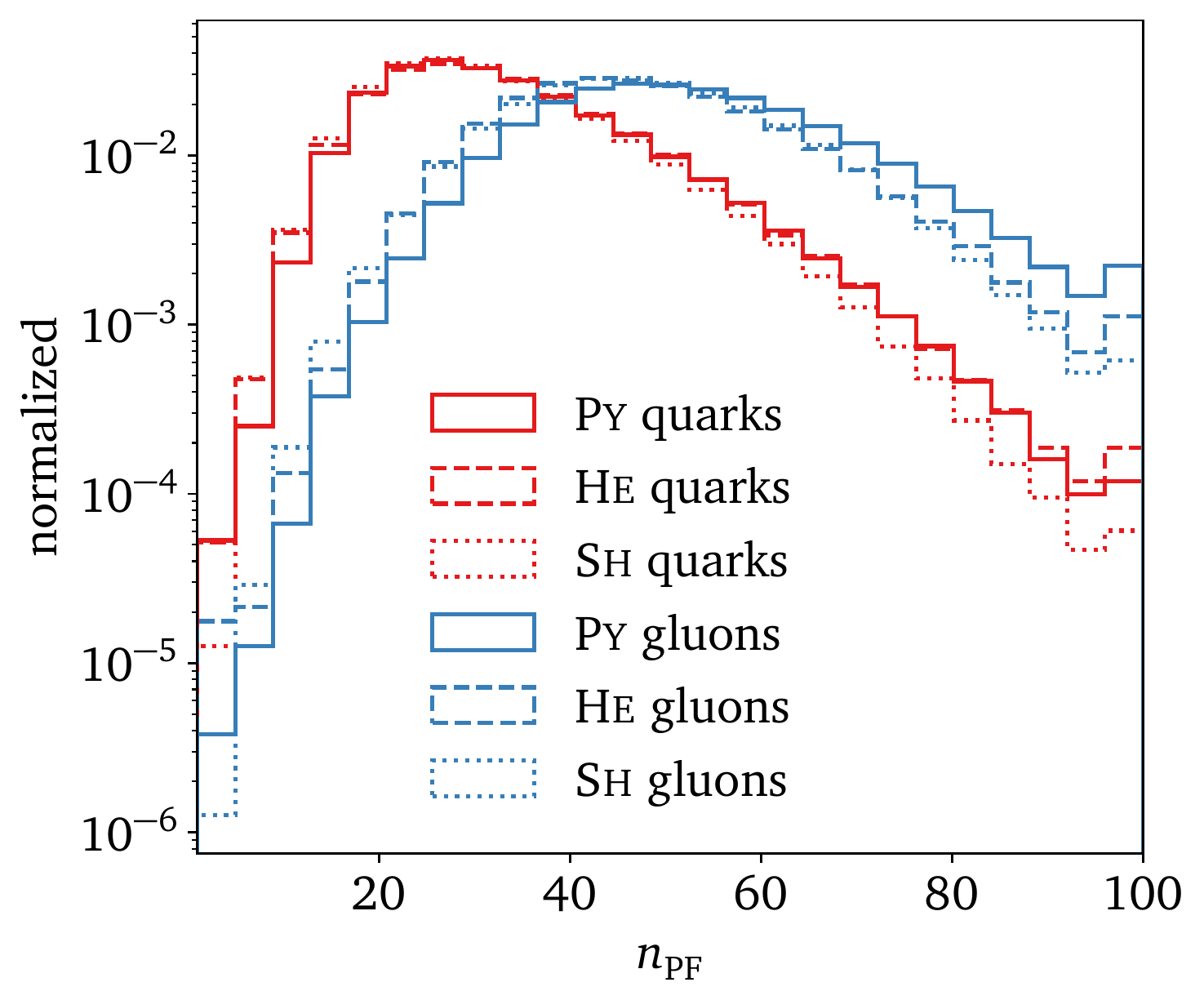}
		%\caption{}\label{fig:}
	\end{subfigure}\hfill
	\begin{subfigure}{0.50\linewidth}
		\centering
		\includegraphics[width=0.94\linewidth]{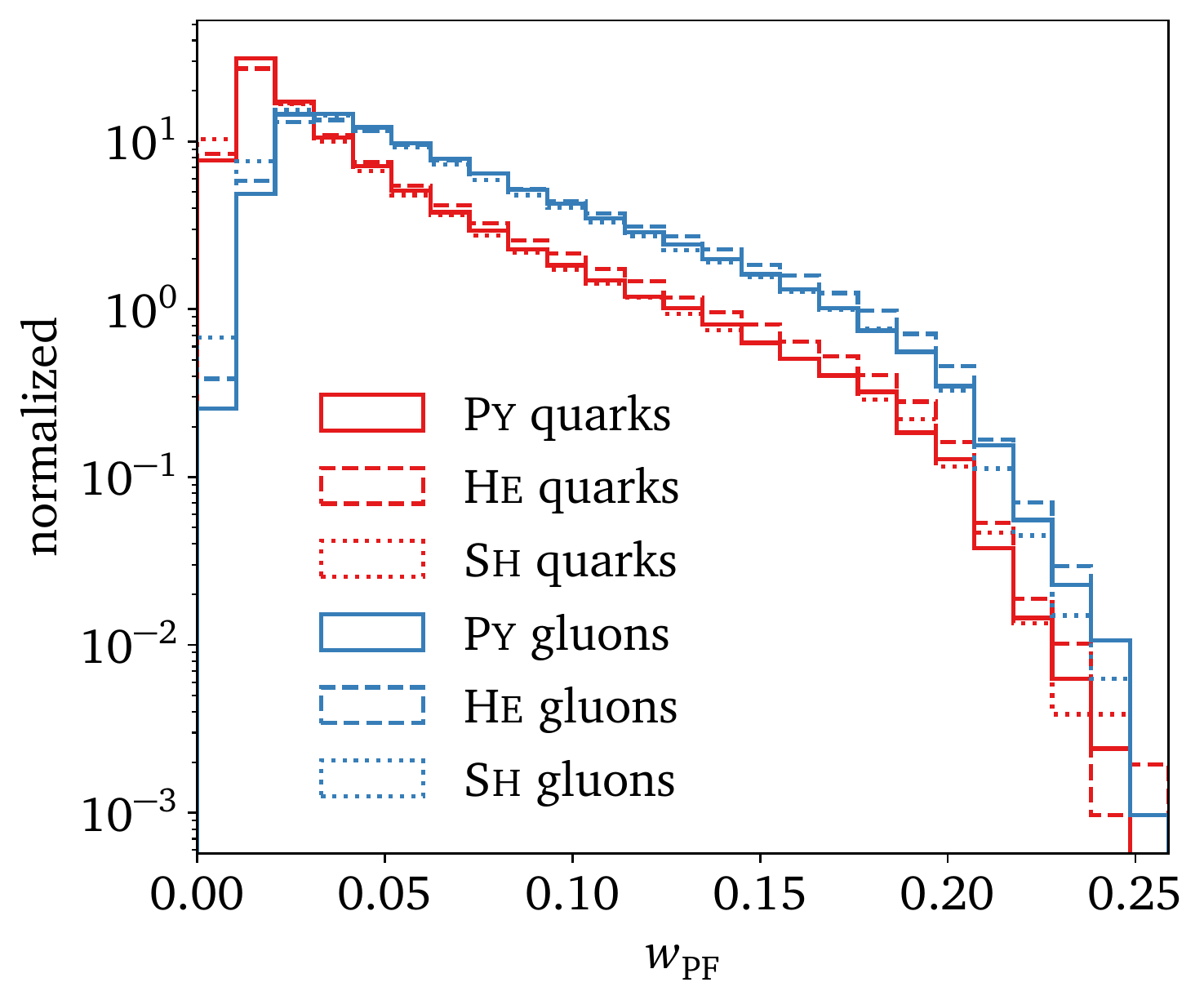}
		%\caption{}\label{fig:}
	\end{subfigure}\newline
	\begin{subfigure}{0.50\linewidth}
		\centering
		\includegraphics[width=0.94\linewidth]{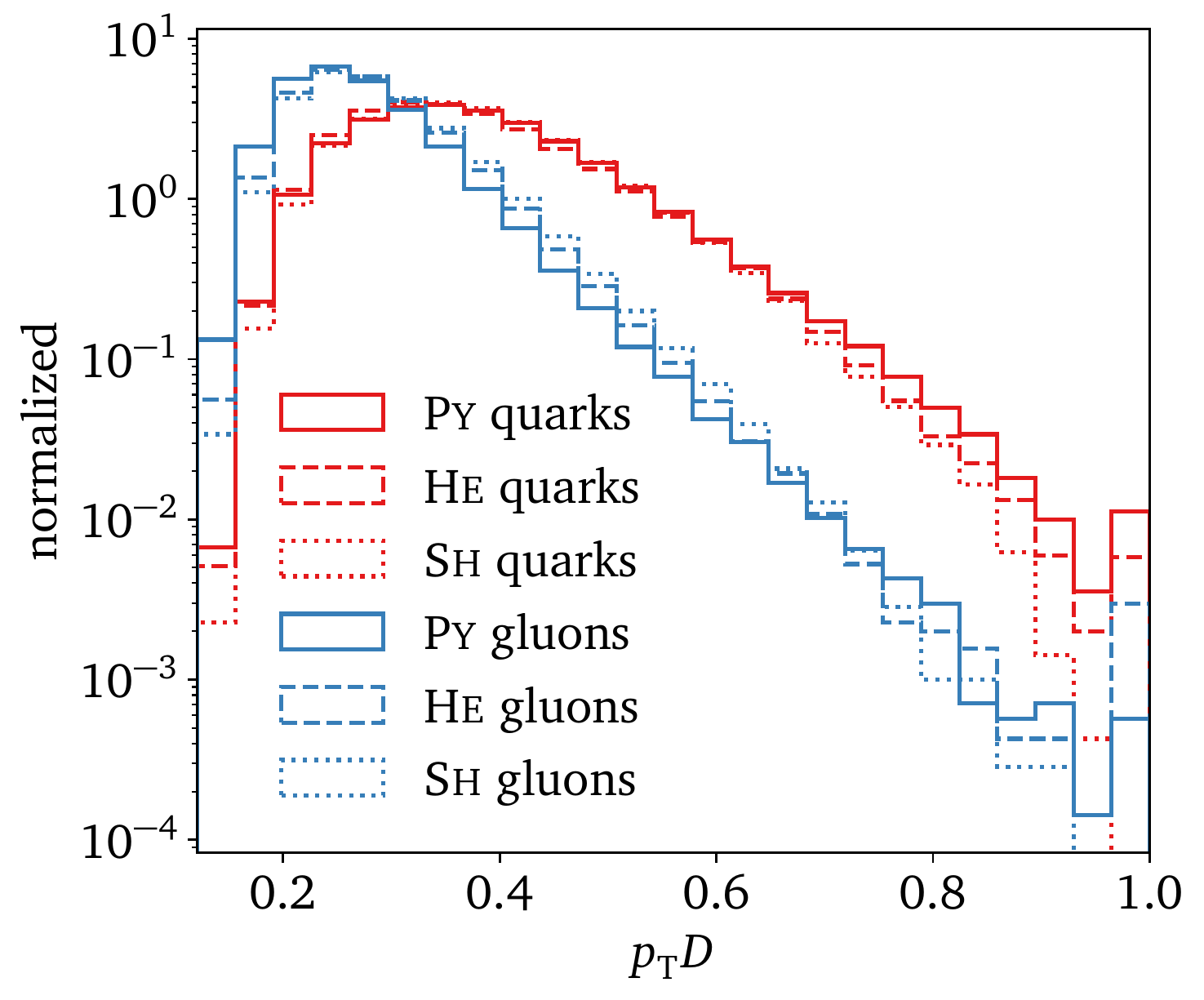}
		%\caption{}\label{fig:}
	\end{subfigure}\hfill
	\begin{subfigure}{0.50\linewidth}
		\centering
		\includegraphics[width=0.94\linewidth]{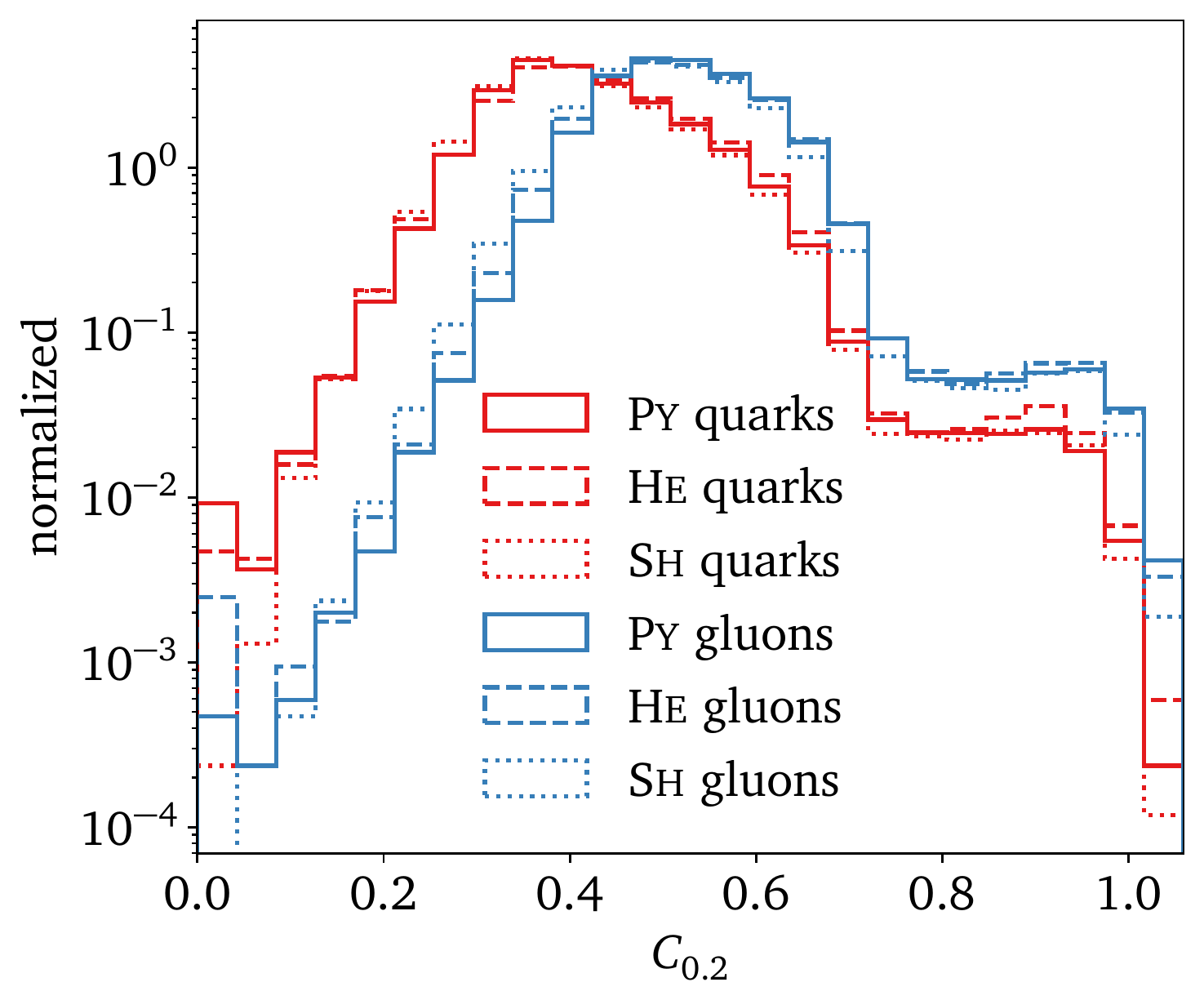}
		%\caption{}\label{fig:}
	\end{subfigure}
    \caption{High-level observables, as defined in \cref{eq:high_level}, for the three different generators, split into quark and gluon jets.}
    \label{fig:qg-histograms}
\end{figure}
%-----------------------------------------------------------

%-----------------------------------------------------------
\begin{table}[b!]
  \centering
  \begin{small}
    \begin{tabular}{l|cccc}
      \toprule
      generator & $\npf$ & $\wpf$ & $\ptd$ & $C_{0.2}$ \\
      \midrule
      \pythia & $2050$ & $3207$ & $4000$ & $1316$ \\
      \sherpa & $1711$ & $3149$ & $3217$ & $1112$ \\
      \herwig & $1326$ & $2910$ & $3406$ & $1128$ \\
      \bottomrule
    \end{tabular}
  \end{small}
  \caption{First Wasserstein distance, or earth mover's distance,
    between quark and gluon distributions for the observables defined
    in \cref{eq:high_level}. We show $200$k quark jets and $200$k gluon jets for each generator.}
  \label{tab:wasserstein-distance}
\end{table}
%-----------------------------------------------------------

%-----------------------------------------------------------
\begin{figure}[t]
    \centering
	\begin{subfigure}{0.50\linewidth}
		\centering
		\includegraphics[width=0.98\linewidth]{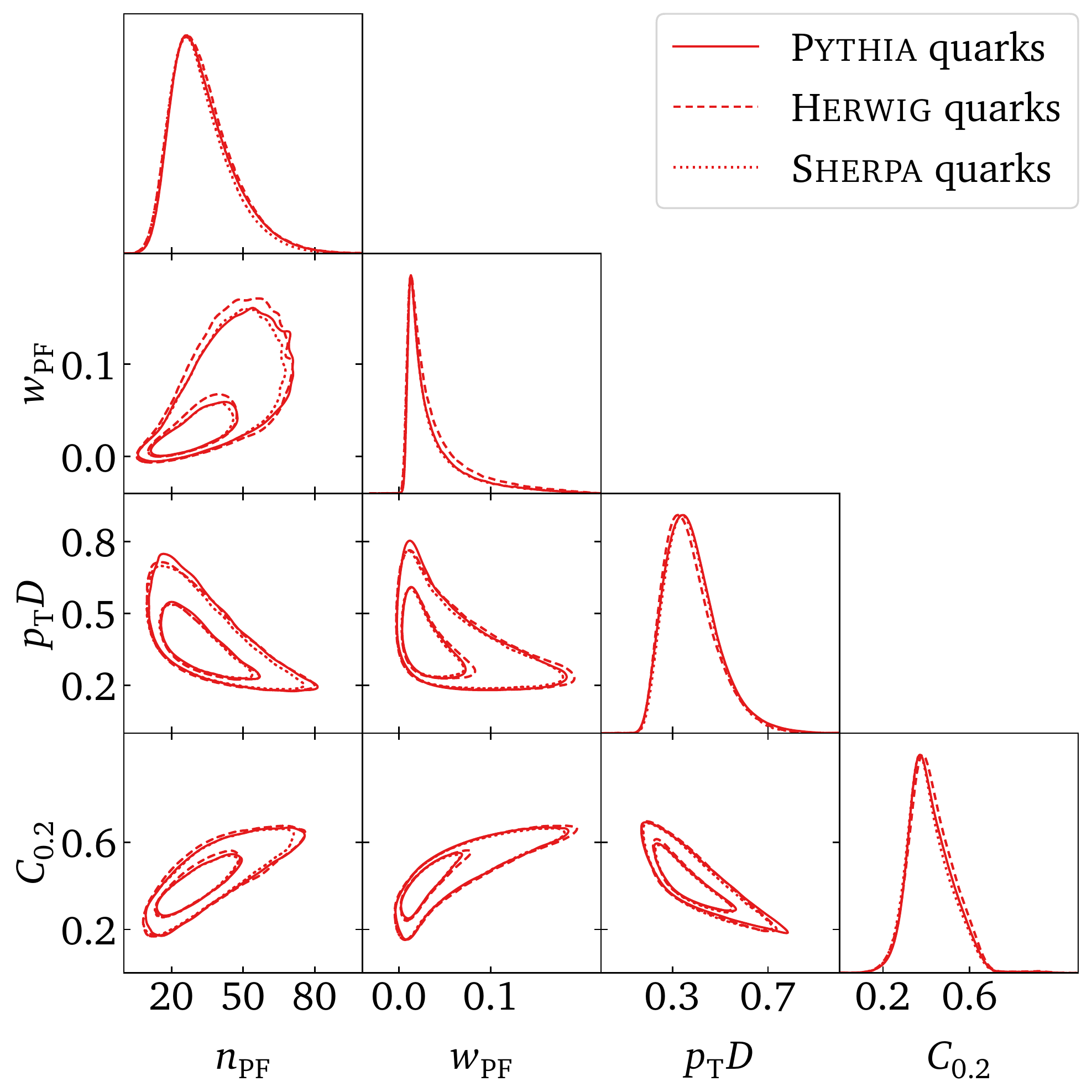}
		%\caption{}\label{fig:}
	\end{subfigure}\hfill
	\begin{subfigure}{0.50\linewidth}
		\centering
		\includegraphics[width=0.98\linewidth]{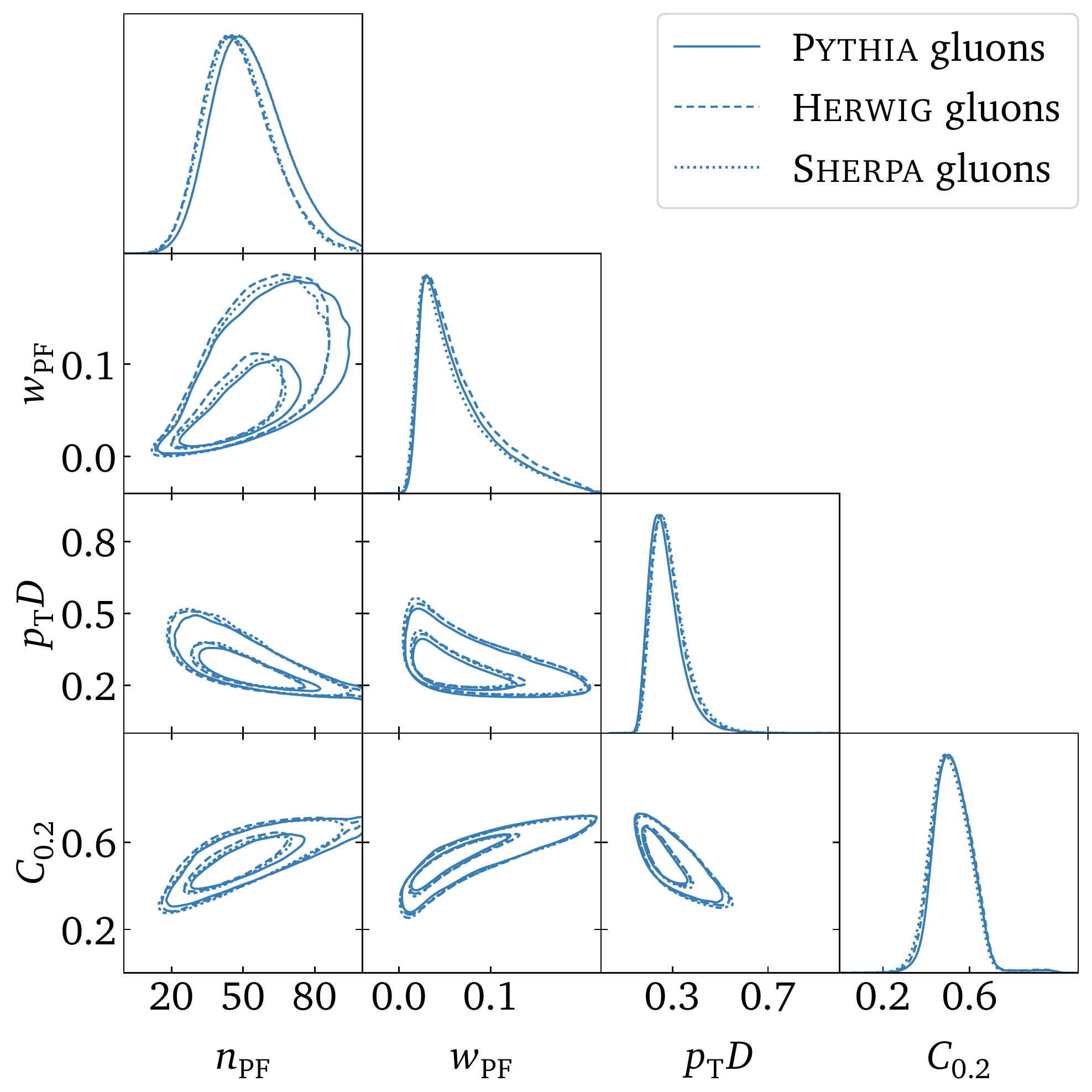}
		%\caption{}\label{fig:}
	\end{subfigure}
    \caption{Correlations between the high-level observables from \cref{eq:high_level}. We show results for the three different generators, split into quark jets (\textit{left}) and gluon jets (\textit{right}).}
	\label{fig:corr}
\end{figure}
%-----------------------------------------------------------

In \cref{fig:qg-histograms} we show these four distributions for
the quark and gluon jets simulated by \pythia, \sherpa, and
\herwig. The biggest difference appears in $\npf$, where the quark
distributions from the three generators are similar, but the gluon
distributions vary significantly. The maximum of the broad peak is the
smallest, $\npf\sim 40$ for the \herwig gluons and the largest, $\npf
\sim \range{45}{50}$ for \pythia gluons. This difference between \herwig
and \pythia jets is not cause by noise or a broader distribution, but
by an actually different prediction for $\npf$.

This difference in $\npf$ vanishes for $\wpf$, indicating that it
comes from infrared and collinear unsafe regions of phase space, and
might become less relevant once we include detector effects. We
emphasize that this does not mean we should expect the shower
algorithms to fail, but that these difference are not easily
computable in perturbative QCD. Similarly, the $\ptd$ distributions
are significantly different for quarks and gluons, combined with a
small shift in the position of the comparably sharp gluon peaks from
the different generators. Finally, the only actual two-constituent
correlation $C_{0.2}$ is also different for quarks and gluons, but
consistent for the different generators. We have studied a range of
additional high-level operators and traced significant deviations
between the gluon jets from the different generators to a strong
correlation with $\npf$.

In \cref{fig:corr} we also show the correlations between the same
observables, for each of the three generators and separated into true
quark and gluon jets. All observables are correlated with the most
powerful $\npf$, but this correlation is not very different for quarks
and for gluons, suggesting that a multi-dimensional analysis will be
dominated by the completely understood shifts in $\npf$.

%-----------------------------------------------------------
\begin{figure}[t]
	\centering
	\begin{subfigure}{0.50\linewidth}
		\centering
		\includegraphics[width=0.95\linewidth]{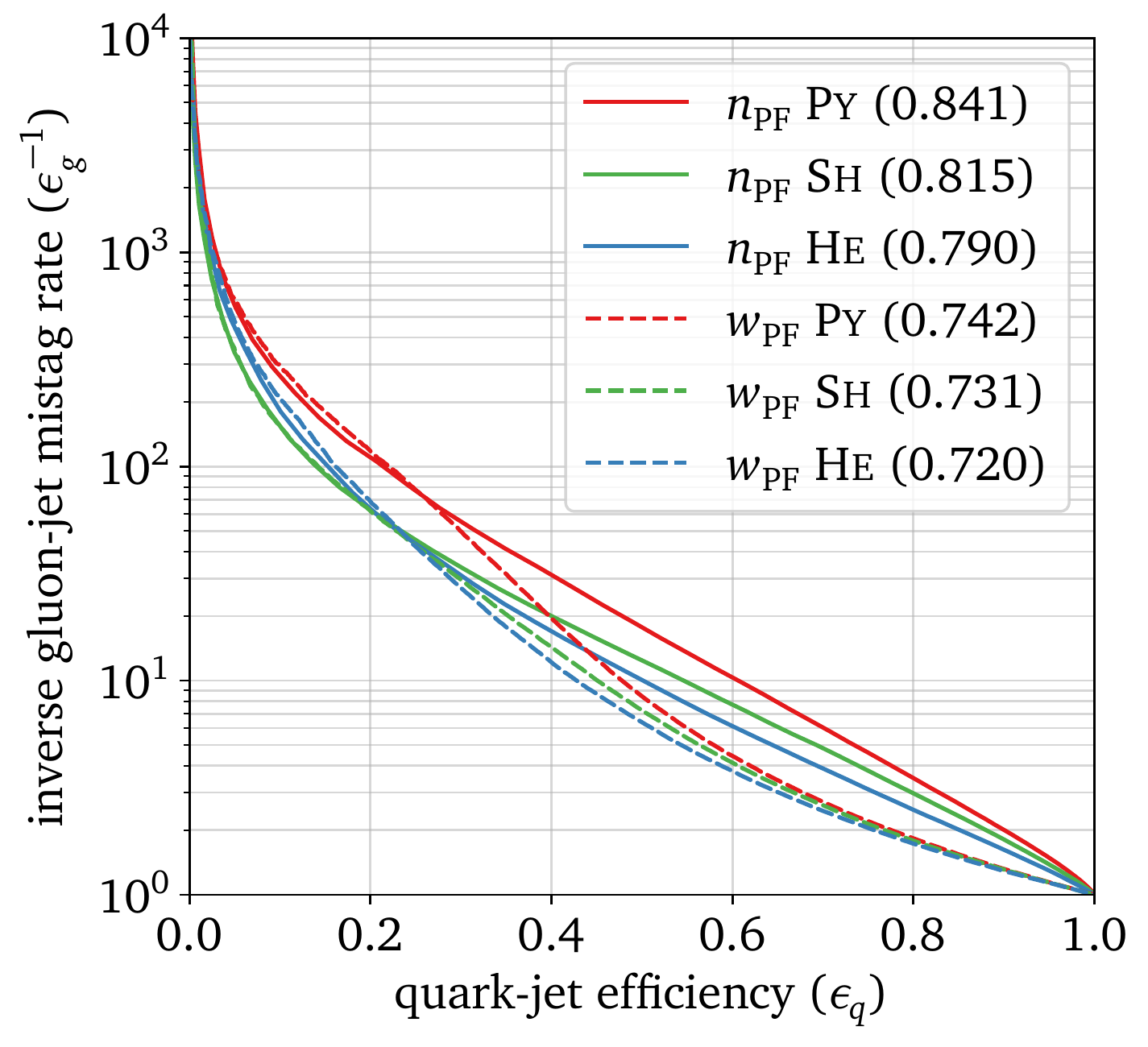}
		%\caption{}\label{fig:}
	\end{subfigure}\hfill
	\begin{subfigure}{0.50\linewidth}
		\centering
		\includegraphics[width=0.95\linewidth]{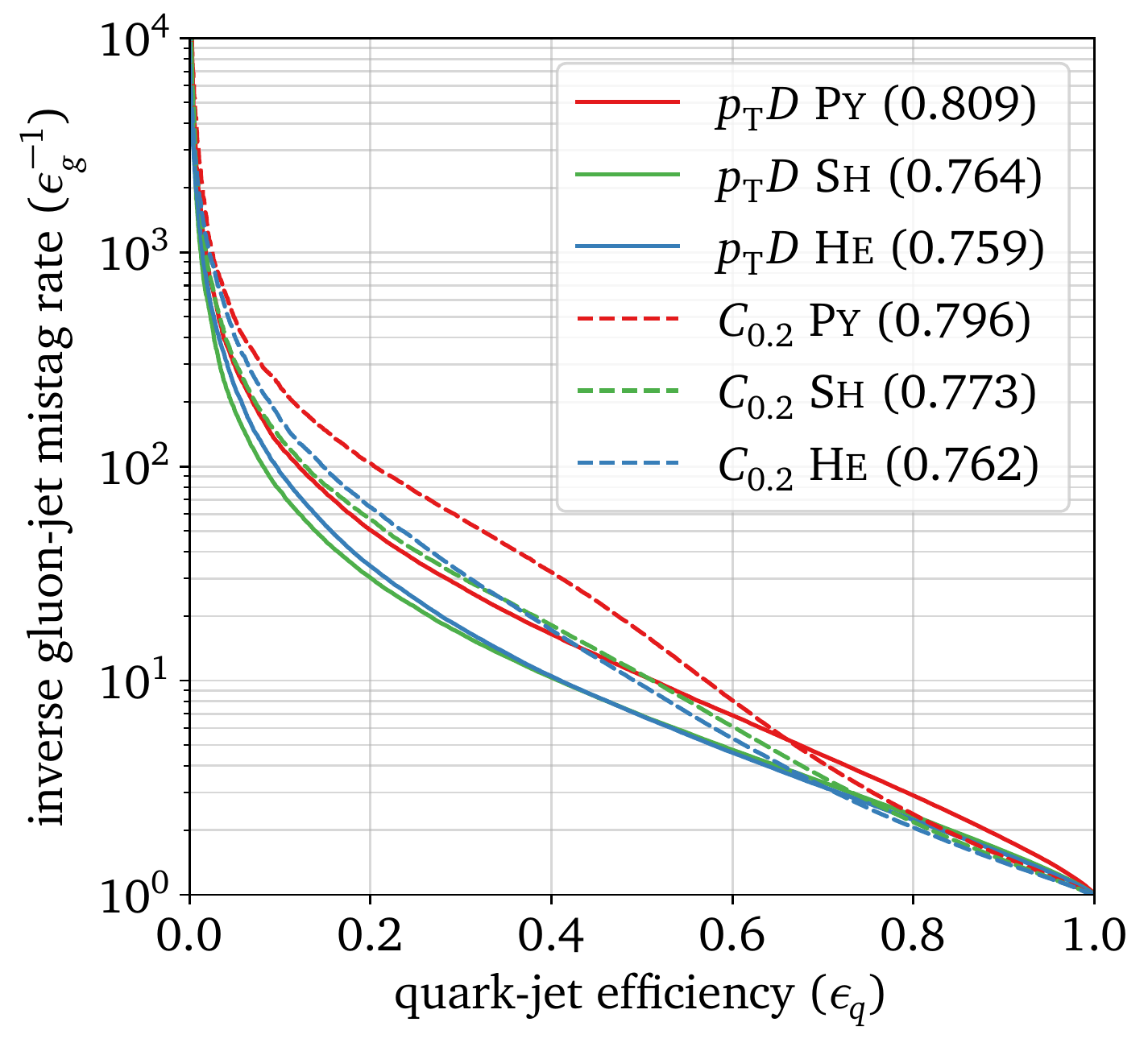}
		%\caption{}\label{fig:}
	\end{subfigure}
    \caption{$\auc$s for quark--gluon discrimination based on
      individual high-level observables of \cref{eq:high_level},
      simulated with \pythia, \sherpa and \herwig.}
	\label{fig:ph-auc}
\end{figure}
%-----------------------------------------------------------

To judge the relevance of the difference in $\npf$ from the different
generators for quark--gluon tagging we can separate quarks from
gluons based on the individual observables given in
\cref{eq:high_level}. We can estimate the power of the individual
distributions using the Wasserstein distances between $200$k quark and
gluon jet histograms, as given in
\cref{tab:wasserstein-distance}. For all observables the \pythia
jets are most easily separated, followed by \sherpa for $\npf$ and
$\wpf$, whereas \herwig predicts a stronger discrimination power for
$\ptd$ than \sherpa. The actual values of the Wasserstein distance for
the different kinematic observables depends on the detailed shape and
does not correlate with the separating power of a kinematic cut. We
show the corresponding ROC curves in \cref{fig:ph-auc}, generated
by choosing such a cut value for each observable. The $\npf$-based and
$\ptd$-based tagging shows a significant degradation when tagging
\herwig jets as compared to the easier-to-separate \pythia quarks and
gluons. This confirms the observation from
\cref{fig:qg-histograms}, where both distributions for \herwig
gluons are further from the common quark distributions than they are
for the \pythia gluons. In contrast, the tagging performance from
$\wpf$ and $C_{0.2}$ is unaffected by the choice of simulation and in
general also much weaker.

To summarize the key result from this simple study --- the most
powerful observables for quark--gluon tagging show a significant shift
in the gluon predictions between \herwig and \pythia. This shift
brings the \herwig gluons closer to quarks.

%%%%%%%%%%%%%%%%%%%%%%%%%%%%%%%%%%%%%%%%%%%%%%%%%%%%%%%%%%%%%%%%%%%%%%%%%%%%%%
\subsection{Bayesian ParticleNet}
\label{sec:specs_net}

To work with a controlled cutting-edge ML-tagger we develop a Bayesian
version of the Particle\-Net\-(-Lite) graph convolutional network
architecture~\cite{Qu:2019gqs} adapted from \tensorflow to \pytorch,
to be able to use our standard Bayesian network. For a detailed
discussion of Bayesian networks we refer to some original Bayesian
network papers~\cite{MacKay:1995bnn,Neal:1995phd,Gal:2016phd} and the
didactic introduction in Ref.~\cite{Plehn:2022ftl}. The \adamw
optimizer~\cite{Kingma:2014vow, Loshchilov:2017wdec}, with a weight
decay of $10^{-4}$, minimizes the usual binary cross-entropy loss
combined with a sigmoid activation function for the classification
task,
\begin{equation}
	\loss_{\mathrm{PN}} = -\frac{1}{M}\sum_{i=1}^{M}
	y_{i}\log f(x_{i})+(1-y_{i})\log\bigl(1-f(x_{i})\bigr) \eqcomma
	\label{eq:loss-bce}
\end{equation}
where $M$ is the mini-batch size, $f(x_{i})\in[0,1]$ the model
prediction for jet $i$, and $y_{i}\in\braces{0,1}$ the jet
truth-label. The two term in the loss lead to a classification of
\begin{equation}
	\begin{split}
		f(x_{i}) \;&\to\; y_{i} = 1 \qqqquad\text{quark signal} \\
		f(x_{i}) \;&\to\; y_{i} = 0 \qqqquad\text{gluon background} \eqperiod
	\end{split}
\end{equation}
We adopt the learning-rate scheduling from Ref.~\cite{Qu:2019gqs}.
The feature input to the ParticleNet are the hardest $100$
jet-constituent particles, specifically
\begin{equation}
	\biggl\lbrace
	\Delta\eta_{k}\:,\;
	\Delta\phi_{k}\:,\;
	\Delta R_{k}\:,\;
	\log p_{\mathrm{T},k}\:,\;
	\log\frac{p_{\mathrm{T},k}}{\ptjet}\:,\;
	\log E_{k}\:,\;
	\log\frac{E_{k}}{E_{\mathrm{jet}}}\:,\;
	\pid_{k}
	\biggr\rbrace \eqcomma
\end{equation}
where the first coordinates are computed relative to the jet axis.
The distance in $\Delta\eta_{k}$ and $\Delta\phi_{k}$ are used to compute
the distances between particles in the first edge convolution
(EdgeConv) block (coordinate input). The PID information includes the
particle charge~\cite{Komiske:2018cqr, Qu:2019gqs}.

While deterministic neural networks adapt a large number of weights to
approximate a training function, Bayesian neural networks (BNNs) learn
distributions of these weights~\cite{Gal:2016phd}.\footnote{In that
  sense there is nothing Bayesian about BNNs, they can just be viewed
  as an extremely efficient way to train network ensembles.} We can
then sample over the weight distributions to produce a central value
and an uncertainty distribution for the network output. In LHC
physics, Bayesian networks can be applied to
classification~\cite{Bollweg:2019skg},
regression~\cite{Kasieczka:2020vlh, Badger:2022hwf}, and generative
networks~\cite{Bellagente:2021yyh, Butter:2021csz, Butter:2022vkj}.
While it is in general possible to separate these uncertainties into
statistical and systematic (stochasticity~\cite{Kasieczka:2020vlh} or
model limitations~\cite{Badger:2022hwf}), we know that our number of
training jets is sufficiently large to only leave us with systematic
uncertainties from the training process.

The Bayesian loss follows from a variational approximation of the
conditional probability for the network parameters. It combines a
likelihood loss with a regularization through a prior for the weight
distributions,
\begin{subequations}\label{eq:loss-bpn}
	\begin{align}
		\loss_{\mathrm{BPN}}
		&=
		-\frac{1}{M}\sum_{i=1}^{M} \log p(y_{i}\vert x_{i}, \omega)
		+\frac{1}{N}\kl\bigl[q_{\mu, \sigma}(\omega), p_{\mu, \sigma}(\omega)\bigr] \\
		&\approx
		-\frac{1}{M}\sum_{i=1}^{M} \log p(y_{i}\vert x_{i}, \omega)
		+\frac{1}{2N}\sum_{\text{weights $\omega_{j}$}}
		\bigl(\mu_{j}^{2}+\sigma_{j}^{2}-\log\sigma_{j}^{2}-1\bigr)
		\eqcomma
	\end{align}
\end{subequations}
where we choose the prior as a Gaussian with mean zero and width one
and use the fact that the resulting weight distributions will become
approximately Gaussian as well, described by $\mu_{j}$ and $\sigma_{j}$.
A change of prior has been shown to not affect the network
output~\cite{Bollweg:2019skg}. As in \cref{eq:loss-bce} $M$
denotes the mini-batch size, and $N$ is the number of training jets.

The parameters $\mu_{j}$ and $\sigma_{j}$ define the model parameters
$\omega_{j}$ of the Bayesian network and need to be trained. In our
case, only the weights in the linear and $2$D-convolutional layers are
extended to Gaussian distributions. The hyperparameters of the
original ParticleNet(-Lite) network and its Bayesian counterparts are
given in \cref{tab:bpn-arch}. We use the same BPN-Lite network for
quark \vs gluon discrimination and for the generator reweighting
which we will introduce in \cref{sec:resilient}.

The performance of the BPN-Lite
quark--gluon classifier is illustrated in the right panel of
\cref{fig:qg-weirdness}. Independent of the competitive $\auc$
values we see that, as before, the network trained and tested on
\pythia performs best, closely followed by the network trained on
\herwig and tested on \pythia. This suggests that the choice of
training sample only has a small effect. In contrast, when we test
networks on \herwig the performance drops significantly, with the
consistent training on \herwig superseding the training on the
alternative \pythia dataset. This hierarchy indicates that, indeed,
\pythia quarks and gluons are easier to separate than the \herwig
quarks and gluons, and that the key features for this classification
are similar for the two generators. We will study this aspect more
closely in the following section.

%-----------------------------------------------------------
\begin{table}[t]
	\centering
	\begin{small}
		\begin{tabular}{l|l}
			\toprule
			hyperparameter & BPN-Lite architecture \\
			\midrule
			number of EdgeConv blocks & $2$ \\
			number of nearest neighbors & $7$ \\
			number of channels for each EdgeConv block &
			$(32,32,32)$, $(64,64,64)$ \\
			channel-wise pooling & average \\
			fully-connected layer & $128$ and ReLU \\
			dropout probability & $0.1$ \\
			number of epochs & $100$ \\
			batch size & $128$ \\
			number of constituents & $100$, with highest $\pt$ \\
			\midrule
			training/validation/testing & $400$k/$100$k/$100$k \\
			signal-to-background ratio & $1.0$ \\
			re-sampling for testing & $80$ times \\
			\bottomrule
		\end{tabular}
	\end{small}
	\caption{Bayesian ParticleNet-Lite (BPN-Lite) architecture and 
	hyperparameters~\cite{Qu:2019gqs}.}
	\label{tab:bpn-arch}
\end{table}
%-----------------------------------------------------------

%%%%%%%%%%%%%%%%%%%%%%%%%%%%%%%%%%%%%%%%%%%%%%%%%%%%%%%%%%%%%%%%%%%%%%%%%%%%%%
\section{Where have all the gluons gone?}
\label{sec:where}

%-----------------------------------------------------------
\begin{figure}[t]
	\begin{minipage}{0.50\linewidth}
		\centering
		\includegraphics[width=0.95\linewidth]{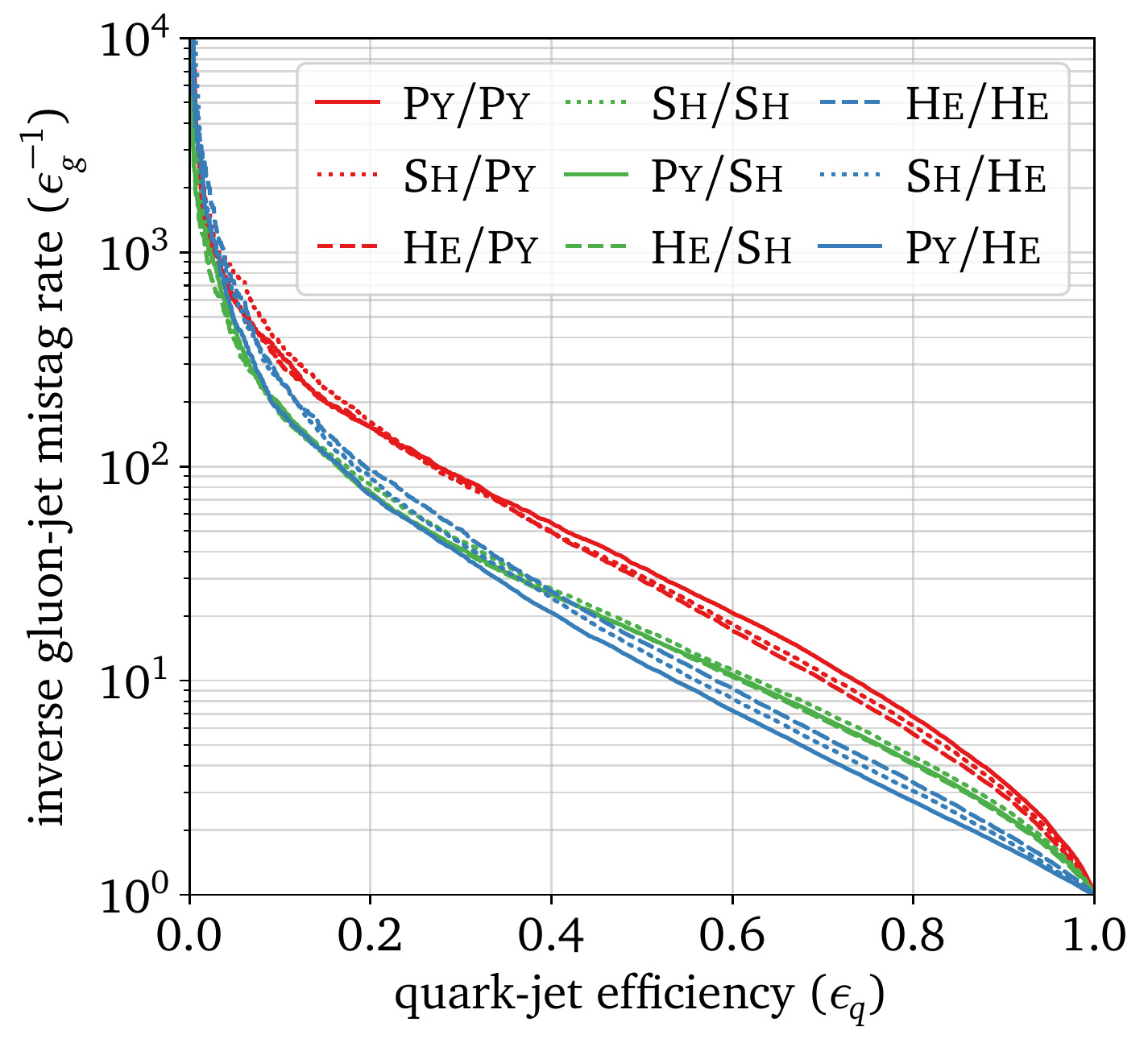}
	\end{minipage}\hfill
	\begin{minipage}{0.50\textwidth}
		\centering
		\begin{small}
			\begin{tabular}{cl|ccc}
				\toprule
				\multicolumn{2}{l|}{$\auc$} & \multicolumn{3}{c}{training} \\
				& & \pythia & \sherpa & \herwig \\
				\midrule
				\parbox[t]{6mm}{\multirow{3}{*}{\rotatebox[origin=c]{90}{testing}}} 
				& \pythia & $0.900$ & $0.893$ & $0.886$ \\
				& \sherpa & $0.856$ & $0.863$ & $0.853$ \\
				& \herwig & $0.804$ & $0.820$ & $0.833$ \\
				\toprule
				\multicolumn{2}{l|}{$\imtafeqg{0.50}$} & \multicolumn{3}{c}{training} \\
				& & \pythia & \sherpa & \herwig \\
				\midrule
				\parbox[t]{6mm}{\multirow{3}{*}{\rotatebox[origin=c]{90}{testing}}} 
				& \pythia & $33.8$ & $30.8$ & $29.4$ \\
				& \sherpa & $16.5$ & $17.4$ & $16.4$ \\
				& \herwig & $12.1$ & $13.8$ & $15.2$ \\
				\bottomrule
			\end{tabular}
		\end{small}
	\end{minipage}
    \caption{\textit{Left:} ROC curves for training and testing on \pythia, \sherpa, and \herwig in different combinations. \textit{Right:} $\auc$ and background rejection performance of the BPN-Lite for quark--gluon tagging, trained and tested on the three different generators.}
    \label{fig:all_auc}
\end{figure}
%-----------------------------------------------------------

Trying to solve the puzzle of quark--gluon taggers trained and tested
on different generators will lead us to the more general question,
namely how to control classification networks trained on one dataset
and tested on another. All combinations of training and testing the
BPN-Lite tagger are illustrated in \cref{fig:all_auc}, with some
of the main results collected in the two tables. Since the original
\pythia and \herwig results form the two extreme poles, we assign the
three datasets to the real-world problem of training a tagger on two
independent training datasets and testing it on independent data as
\begin{itemize}
    \item labelled training dataset~$1$: \pythia,
    \item labelled training dataset~$2$: \herwig,
    \item independent test dataset: \sherpa.
\end{itemize}
We will start by comparing different trainings on the labelled \pythia
and \herwig datasets, as motivated by \cref{fig:qg-weirdness} and
eventually add \sherpa results as an independent test, in the sense of
actual data analyzed by the tagger.

%%%%%%%%%%%%%%%%%%%%%%%%%%%%%%%%%%%%%%%%%%%%%%%%%%%%%%%%%%%%%%%%%%%%%%%%%%%%%%
\subsubsection*{Performance comparison}

\begin{figure}[t]
	\centering
	\begin{subfigure}{0.50\linewidth}
		\centering
		\includegraphics[width=0.94\linewidth]{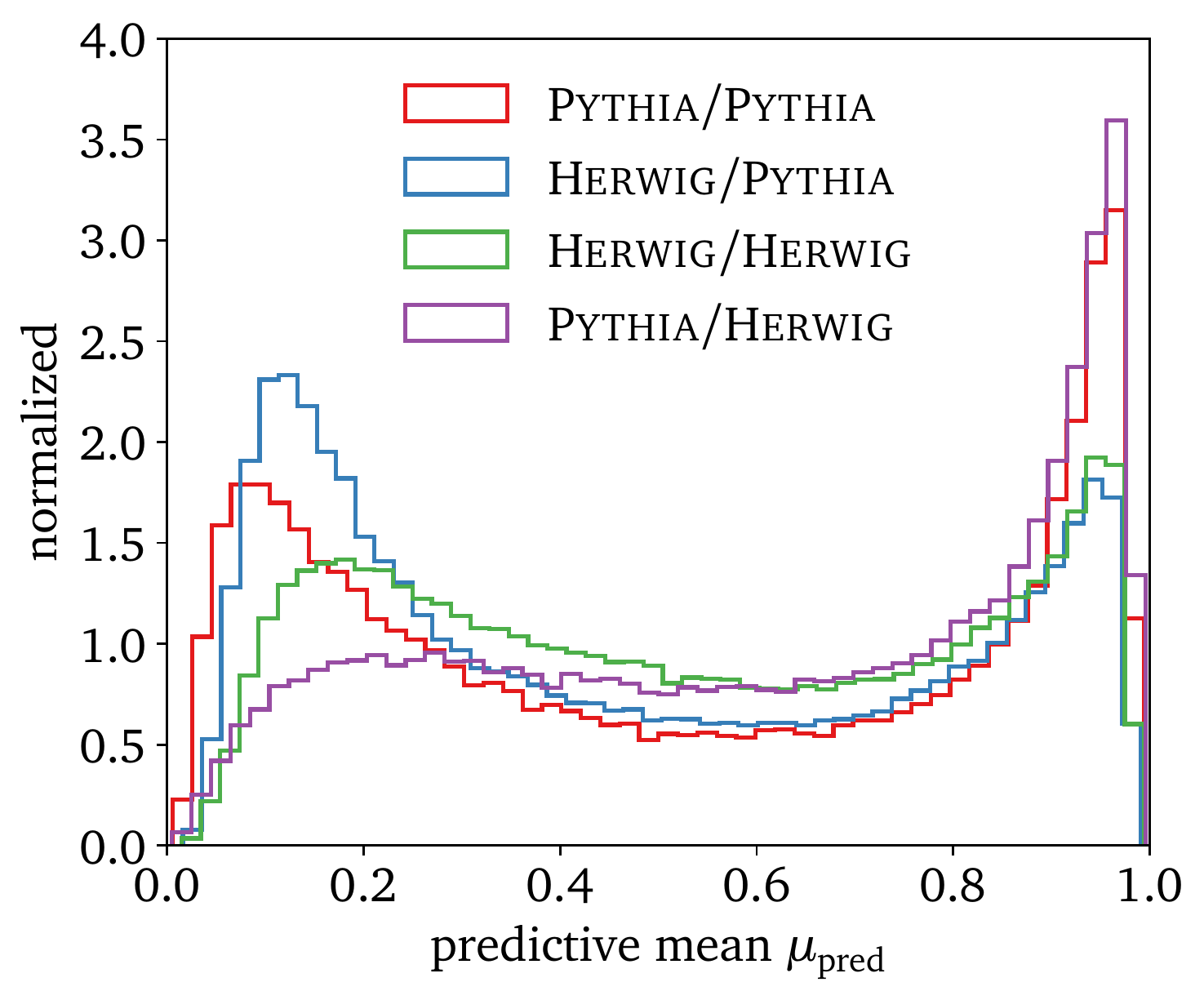}
		%\caption{}\label{fig:}
	\end{subfigure}\hfill
	\begin{subfigure}{0.50\linewidth}
		\centering
		\includegraphics[width=0.94\linewidth]{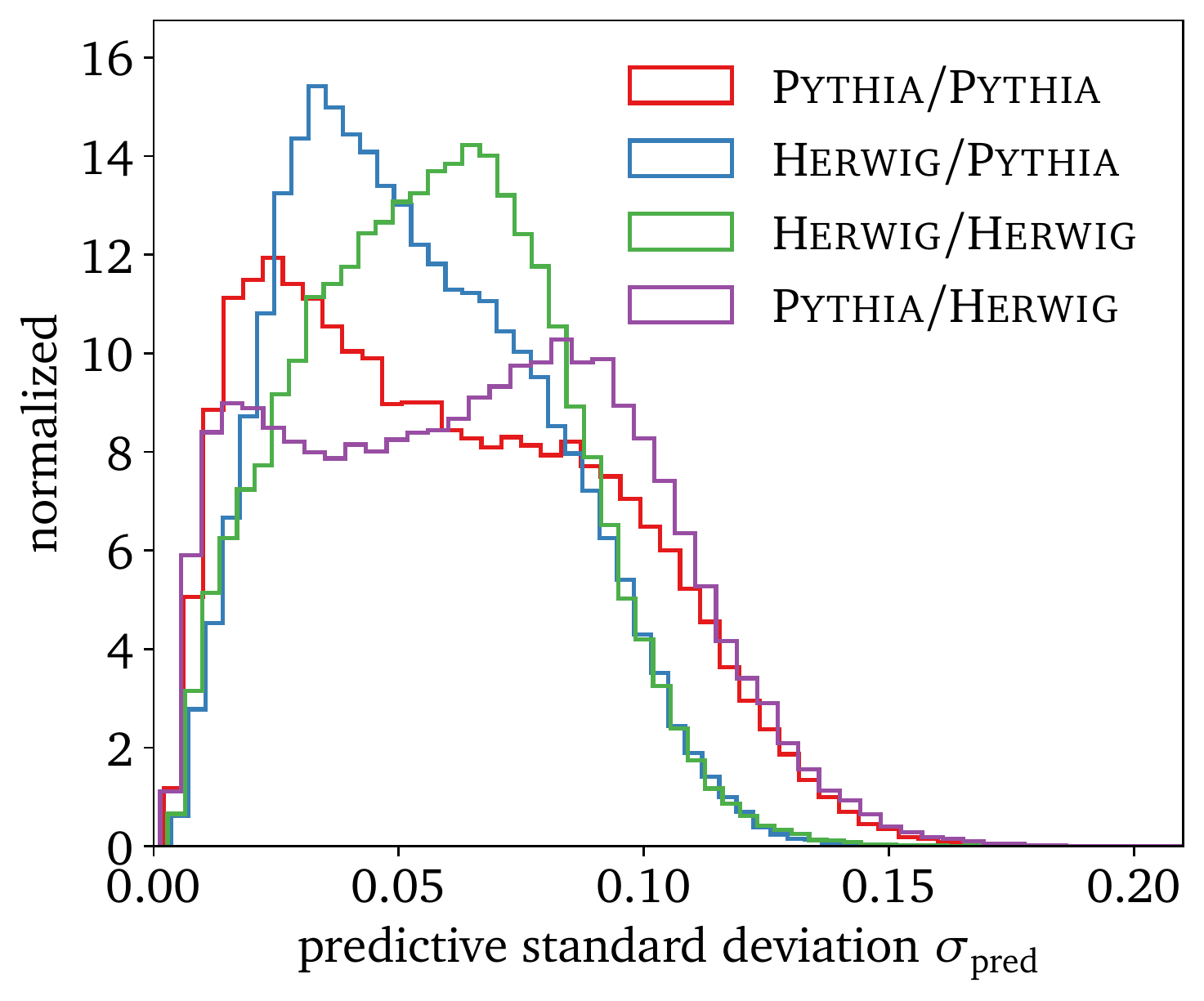}
		%\caption{}\label{fig:}
	\end{subfigure}\newline
	\begin{subfigure}{0.25\linewidth}
		\centering
		\includegraphics[width=0.98\linewidth]{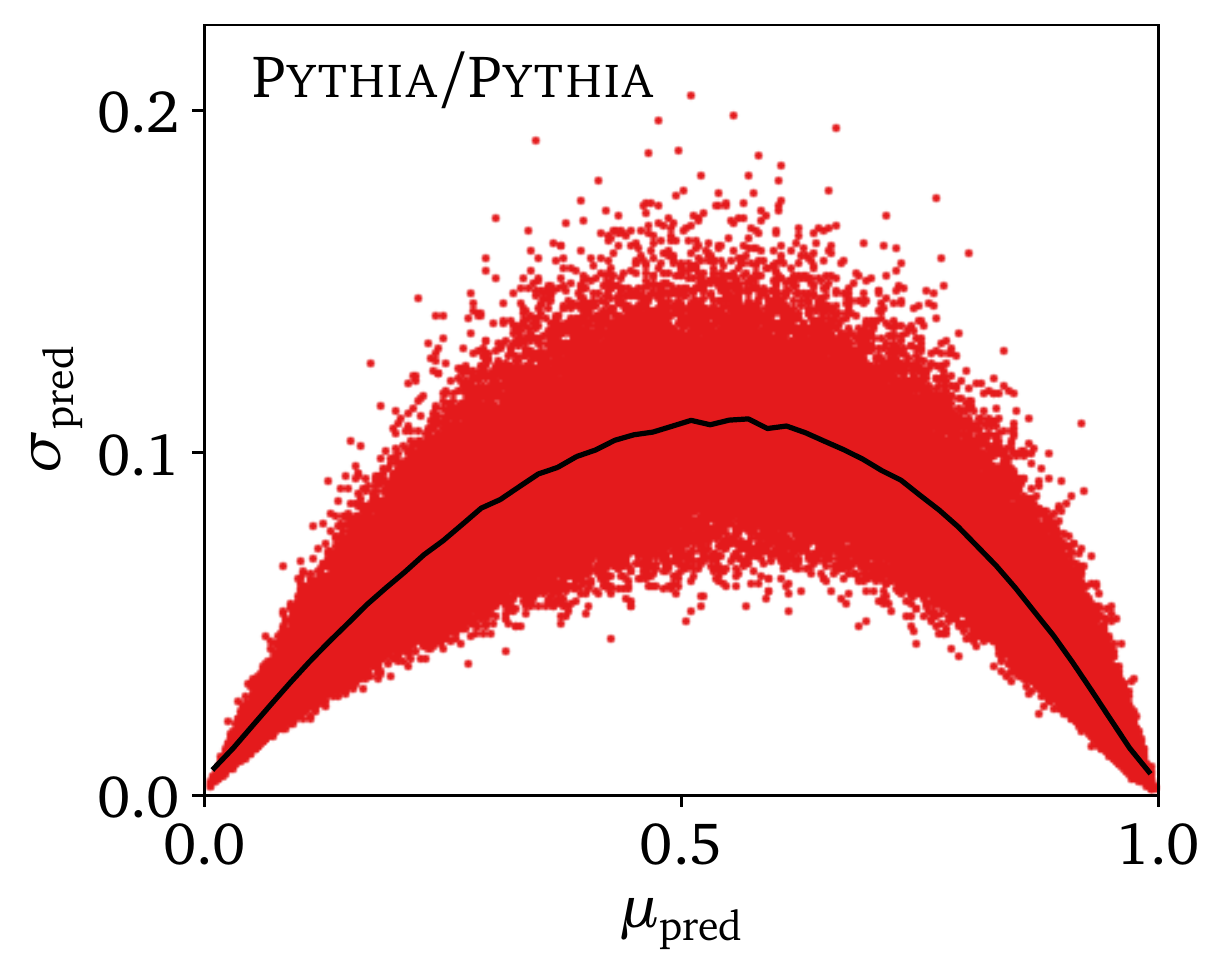}
		%\caption{}\label{fig:}
	\end{subfigure}\hfill
	\begin{subfigure}{0.25\linewidth}
		\centering
		\includegraphics[width=0.98\linewidth]{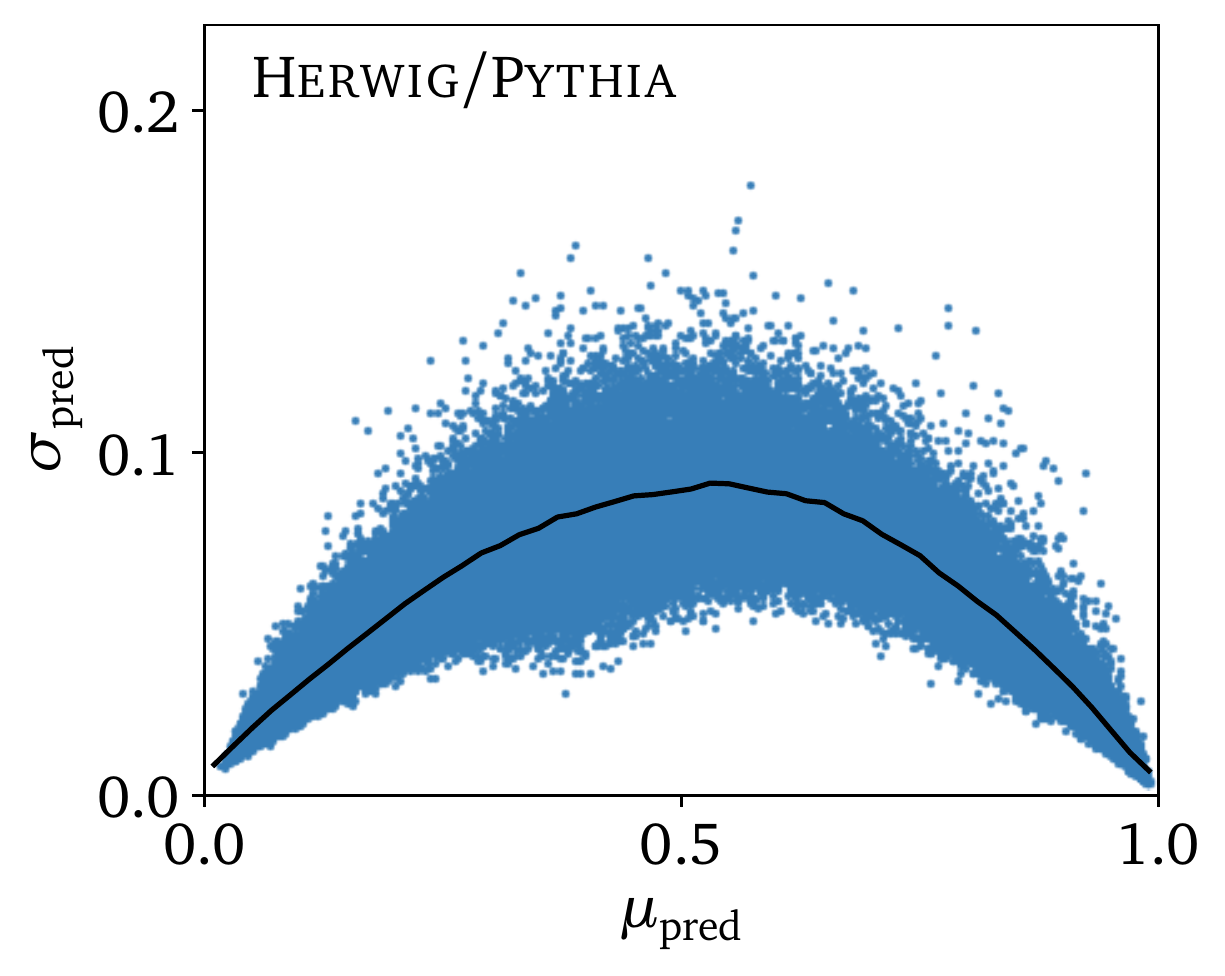}
		%\caption{}\label{fig:}
	\end{subfigure}\hfill
	\begin{subfigure}{0.25\linewidth}
		\centering
		\includegraphics[width=0.98\linewidth]{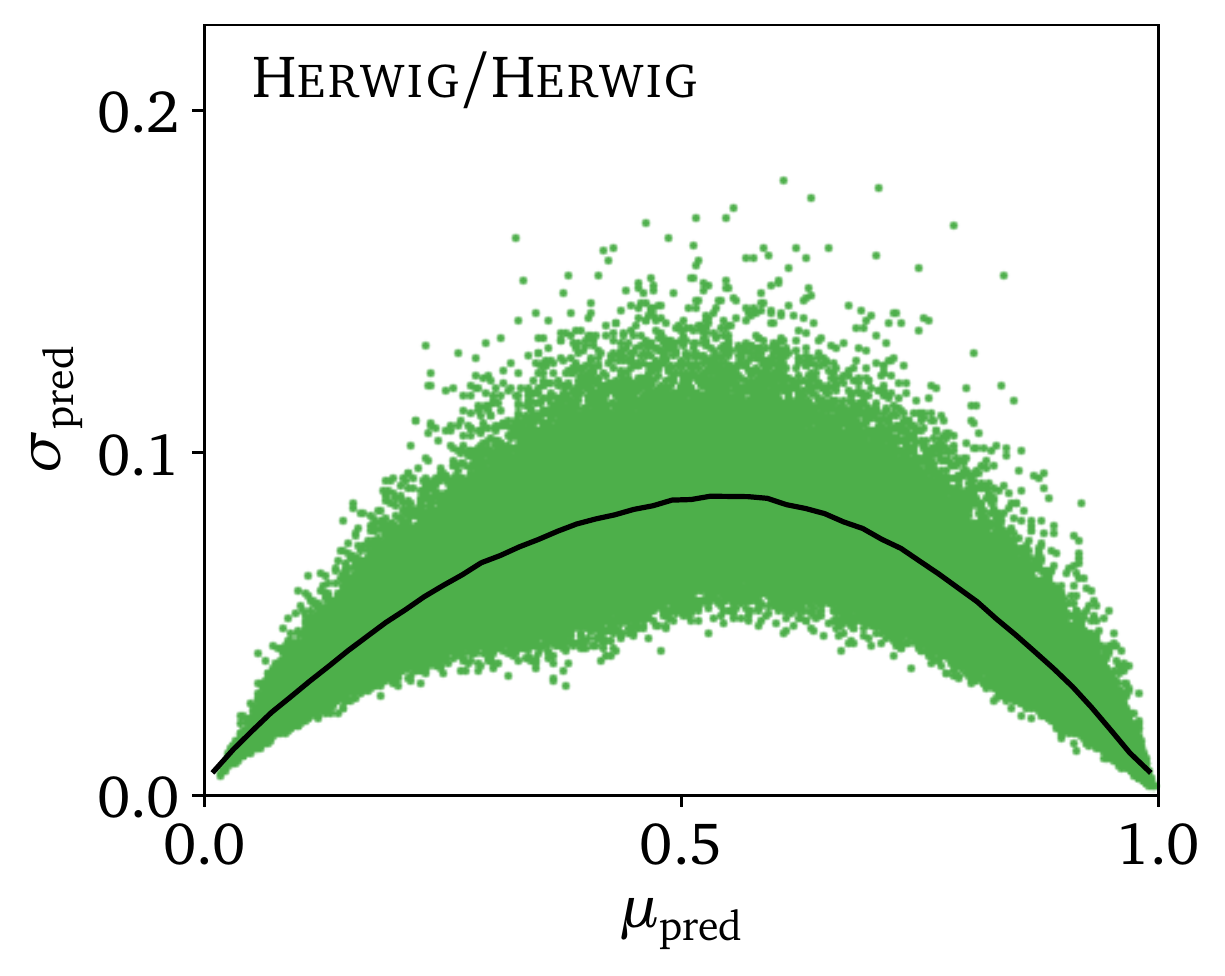}
		%\caption{}\label{fig:}
	\end{subfigure}\hfill
	\begin{subfigure}{0.25\linewidth}
		\centering
		\includegraphics[width=0.98\linewidth]{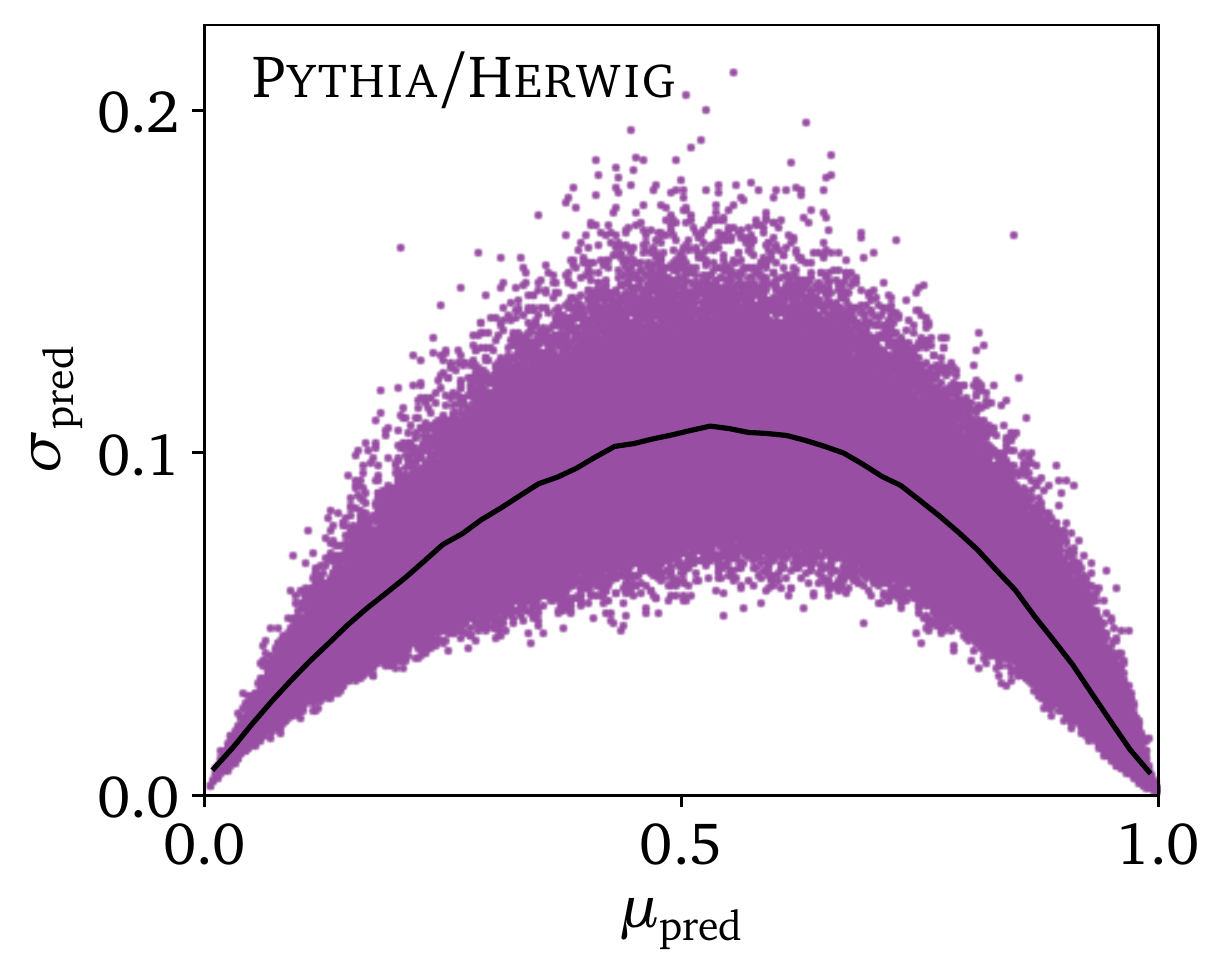}
		%\caption{}\label{fig:}
    \end{subfigure}
    \caption{Predictive means ($\mup=0$ for gluons, $\mup=1$ for quarks) and standard deviations from the BPN-Lite tagger trained and tested on \pythia and \herwig in different combinations. The lower panels illustrate a stochastic pattern around the correlation of \cref{eq:mu_sigma}.}
  \label{fig:bpn-mean-std}
\end{figure}

The Bayesian nature of the BPN-Lite tagger comes with two pieces of
information, which allow us to understand the network training. First,
the Bayesian tagger provides a per-jet uncertainty $\sip(x_{i})$ on the
classification output $\mup(x_{i})\in[0,1]$. This means we can
separate jets for which the network training leads to a confident
classification from jets where the training provides less
information. Second, the final sigmoid layer of the classification
network leads to a correlation of $\mup$ and $\sip$, namely
\begin{equation}
	\sip(x_{i})\:\propto\:\mup(x_{i})\,\bigl[1-\mup(x_{i})\bigr] \eqperiod
	\label{eq:mu_sigma}
\end{equation}
This inverse parabola correlation is a feature of the network
structure and has to be present in the Bayesian tagging output, its
absence points to a stability issue in the networks training. In
\cref{fig:bpn-mean-std} we show the $\mup$- and
$\sip$-distributions for \pythia and \herwig test datasets, after
consistently training the networks on \pythia and \herwig. Already the
$\mup$-distributions shows three major issues:
\begin{enumerate}
    \item While the tagging of quarks \vs gluons is never symmetric,
    training and testing on \pythia indicates some gluons confidently
    identified as gluons $\mup(x_{i})\to 0$.
    \item Training and testing on \herwig hardly ever allows the network
    to confidently identify gluons with $\mup\lesssim 0.1$.
    \item Training on \pythia and testing on \herwig identifies at least some gluons with as small $\sip$ as training and testing on \pythia.
\end{enumerate}
Looking at the $\sip$-distribution, the results from training on
\pythia look as expected as long as we test on \pythia jets, but
tested on \herwig a slight shoulder around $\sip\sim 0.07$ develops
into a second peak. This peak corresponds to jets or phase-space
configurations where the \pythia training does not allow for a
confident application to \herwig jets. Second, the general uncertainty
after training on \herwig jets peaks at larger $\sip$, indicating that
the network faces difficulties to extract the relevant features for
the tagging, but also drops off at smaller $\sip$ values than the
\pythia trainings. This reflects the problem with the single main
feature $\npf$, as expected from our discussion in \cref{sec:specs_qg}.

%-----------------------------------------------------------
%-----------------------------------------------------------

Finally, the four lower panels in \cref{fig:bpn-mean-std} show the
per-jet correlation of the predictive means and standard
deviations. Again confirming our suspicions from
\cref{sec:specs_qg} that training on \herwig jets is not
completely stable, leading to slight irregularities of the scattering
pattern around the inverse parabola predicted by
\cref{eq:mu_sigma}.

%%%%%%%%%%%%%%%%%%%%%%%%%%%%%%%%%%%%%%%%%%%%%%%%%%%%%%%%%%%%%%%%%%%%%%%%%%%%%%
\subsubsection*{High-level observables}

%-----------------------------------------------------------
\begin{figure}[t]
	\centering
	\begin{subfigure}{0.50\linewidth}
		\centering
		\includegraphics[width=0.94\linewidth]{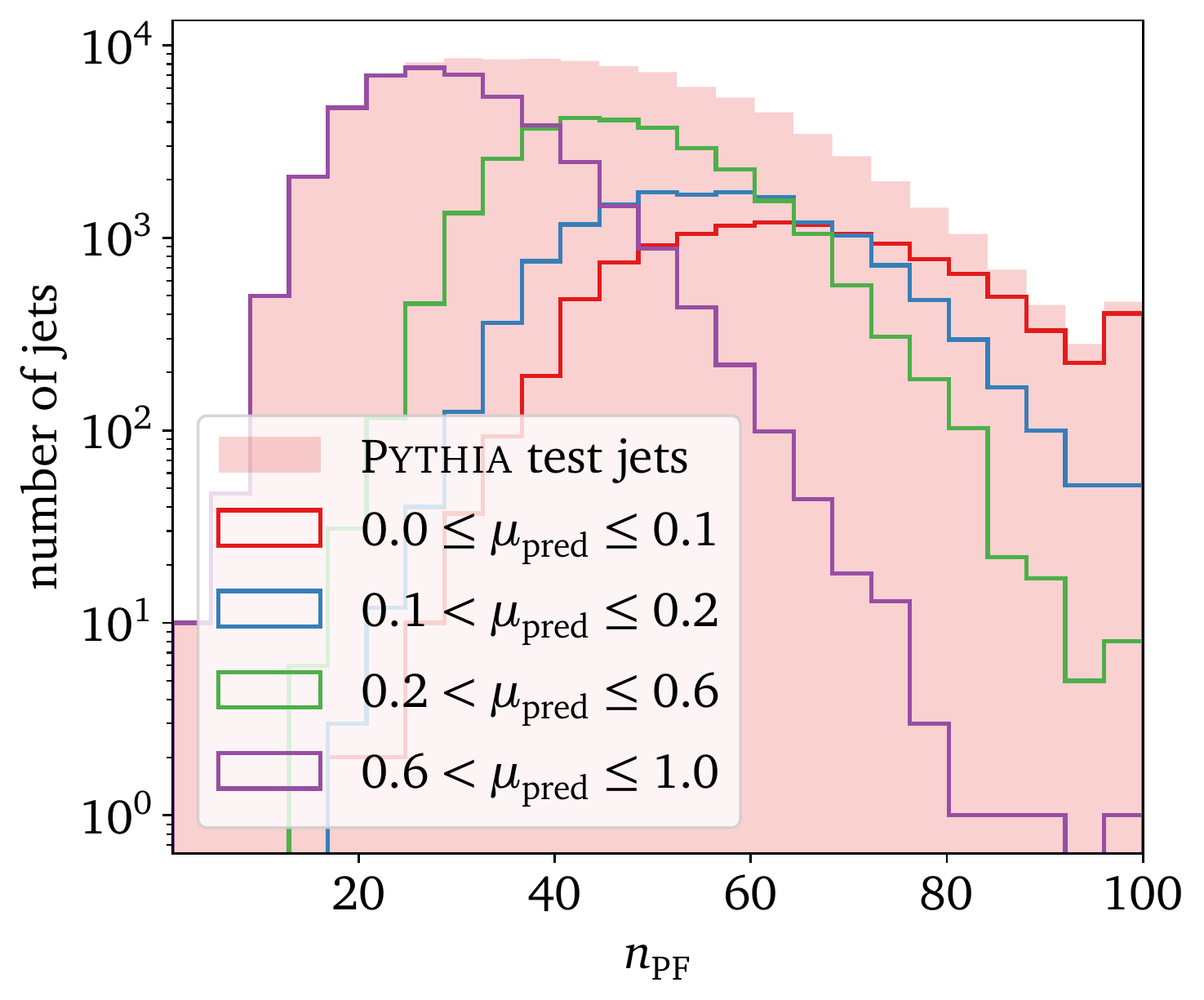}
		%\caption{}\label{fig:}
	\end{subfigure}\hfill
	\begin{subfigure}{0.50\linewidth}
		\centering
		\includegraphics[width=0.94\linewidth]{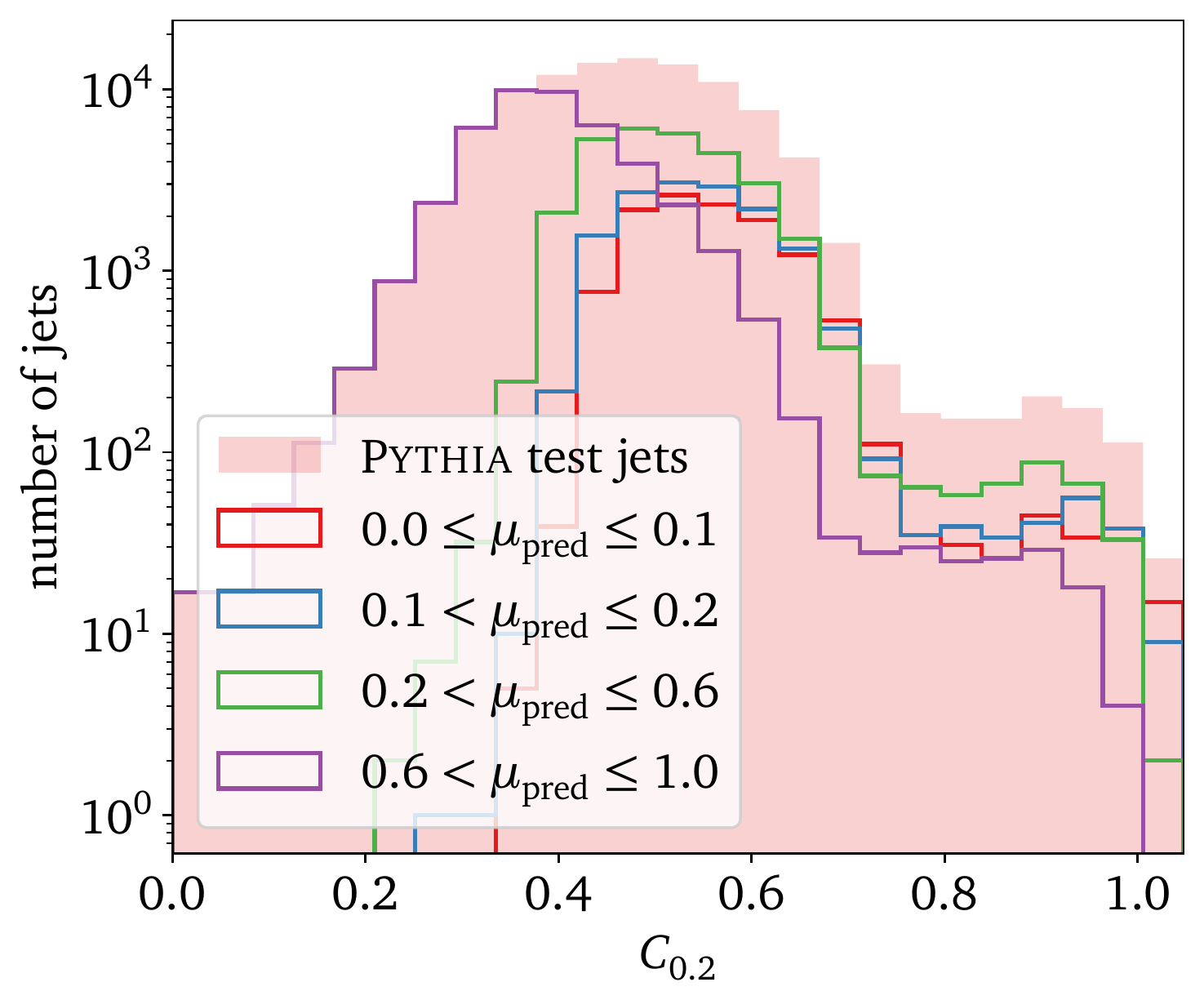}
		%\caption{}\label{fig:}
	\end{subfigure}\newline
	\begin{subfigure}{0.50\linewidth}
		\centering
		\includegraphics[width=0.94\linewidth]{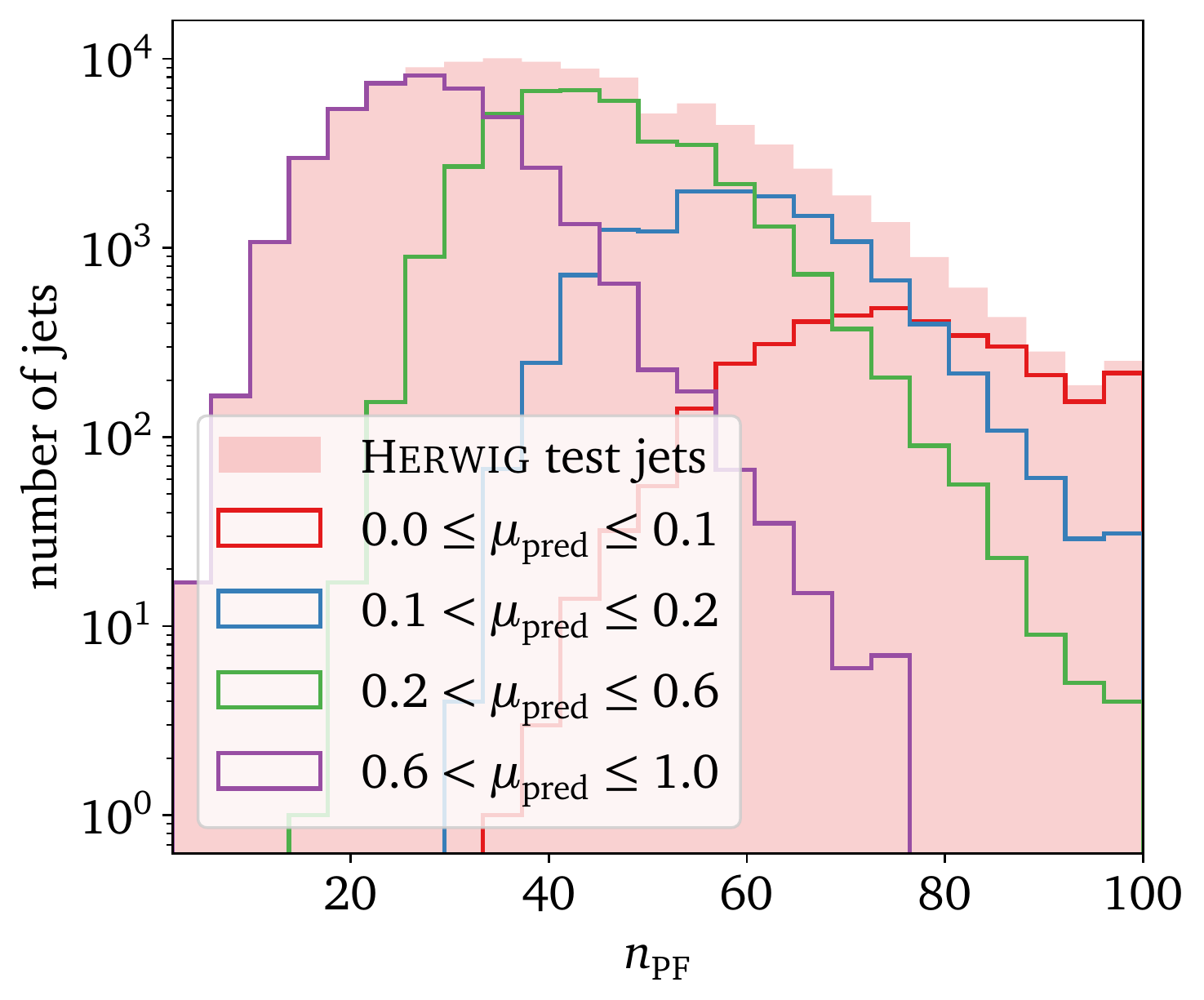}
		%\caption{}\label{fig:}
	\end{subfigure}\hfill
	\begin{subfigure}{0.50\linewidth}
		\centering
		\includegraphics[width=0.94\linewidth]{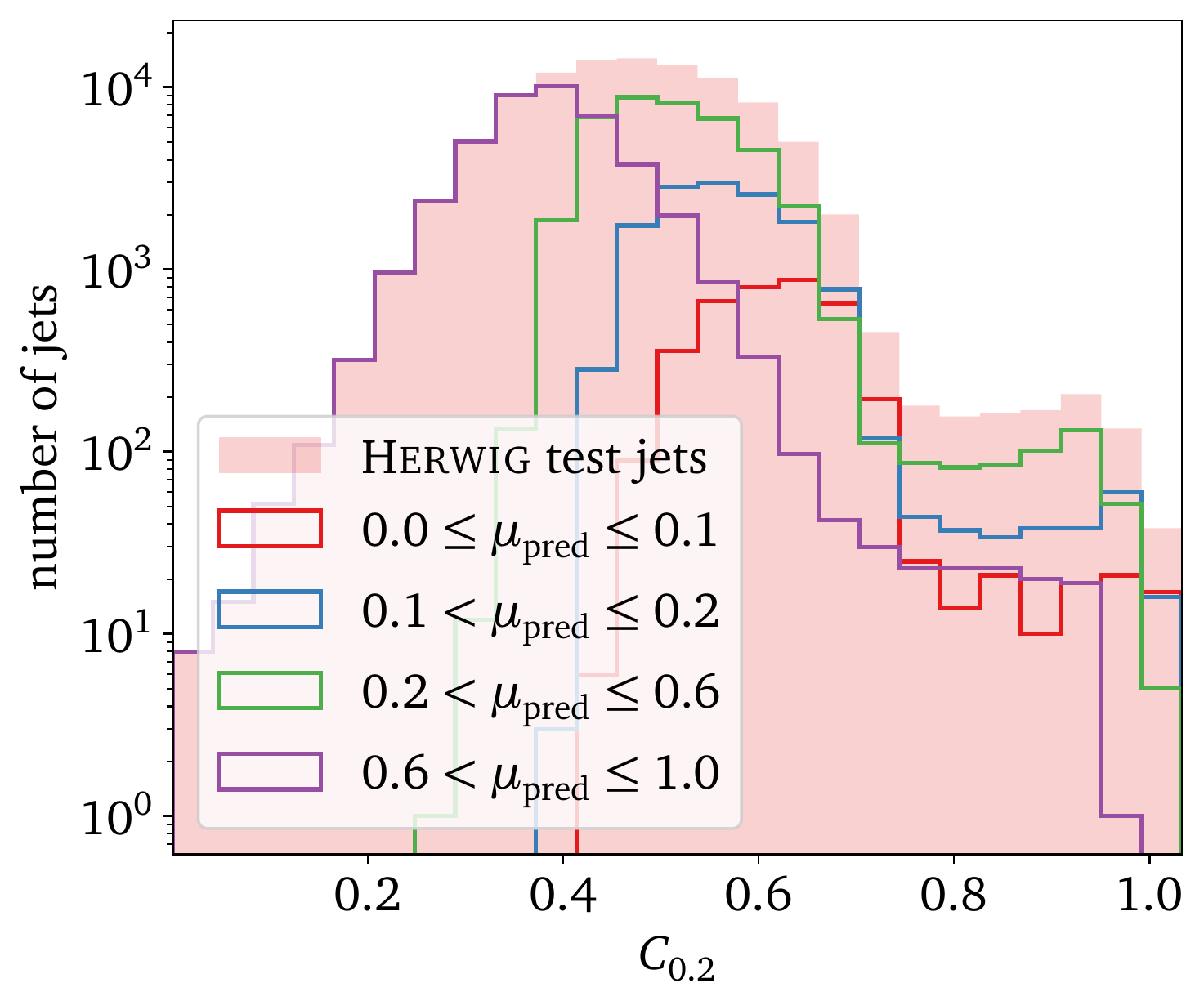}
		%\caption{}\label{fig:}
	\end{subfigure}
    \caption{Kinematic distributions defined in \cref{eq:high_level}. We train the BPN-Lite tagger consistently trained and tested on \pythia (\textit{upper}) and on \herwig (\textit{lower}). The histograms are normalized such that they reflect the fractions of jets in the respective slices in $\mup$, extracted from consistent testing.}
  \label{fig:obs-mu-slices}
\end{figure}
%-----------------------------------------------------------

We can trace back the problems with the performance and stability of
the \herwig training to the high-level observables of
\cref{eq:high_level}. In \cref{fig:obs-mu-slices} we show two
of the most interesting kinematic variables in slices of $\mup$, the
probabilistic output of BPN-Lite. We know already that $\npf$ is the
leading discriminating feature separating quarks from gluons, while
$C_{0.2}$ is the only actual correlator amongst the standard
high-level observables. In the upper panels we show \pythia jets, in
the lower panels \herwig jets. The slices are bases on consistent
training and testing on the two samples. For $\mup>0.6$ the two
distributions agree, as expected for correctly identified quarks.

While the two $\npf$-distributions are very similar for correctly
identified quark-like jets with $\mup>0.6$, differences appear towards the
gluon regime and become quite dramatic for correctly identified gluons
with $\mup<0.1$. Requiring increasingly small $\mup$ values for
more and more confidently identified gluons, the fraction of jets
remaining in these slices from the \pythia sample is much larger than
it is for the \herwig sample. While for \pythia jets values $\npf>60$
indicate confidently identified gluons, \herwig gluons are harder
to identify and typically require $\npf>70$ to lead to the rare
occurrence of $\mup<0.1$. In the right panels we show the correlator
$C_{0.2}$. While the main difference is the number of jets in the
individual slices, we also see that the secondary maximum around
$C_{0.2}>0.8$ is predominantly, but not exclusively populated by
gluon jets.

%%%%%%%%%%%%%%%%%%%%%%%%%%%%%%%%%%%%%%%%%%%%%%%%%%%%%%%%%%%%%%%%%%%%%%%%%%%%%%
\subsubsection*{Predictive uncertainties}

%-----------------------------------------------------------
\begin{figure}[t]
    \centering
    \includegraphics[width=0.495\linewidth]{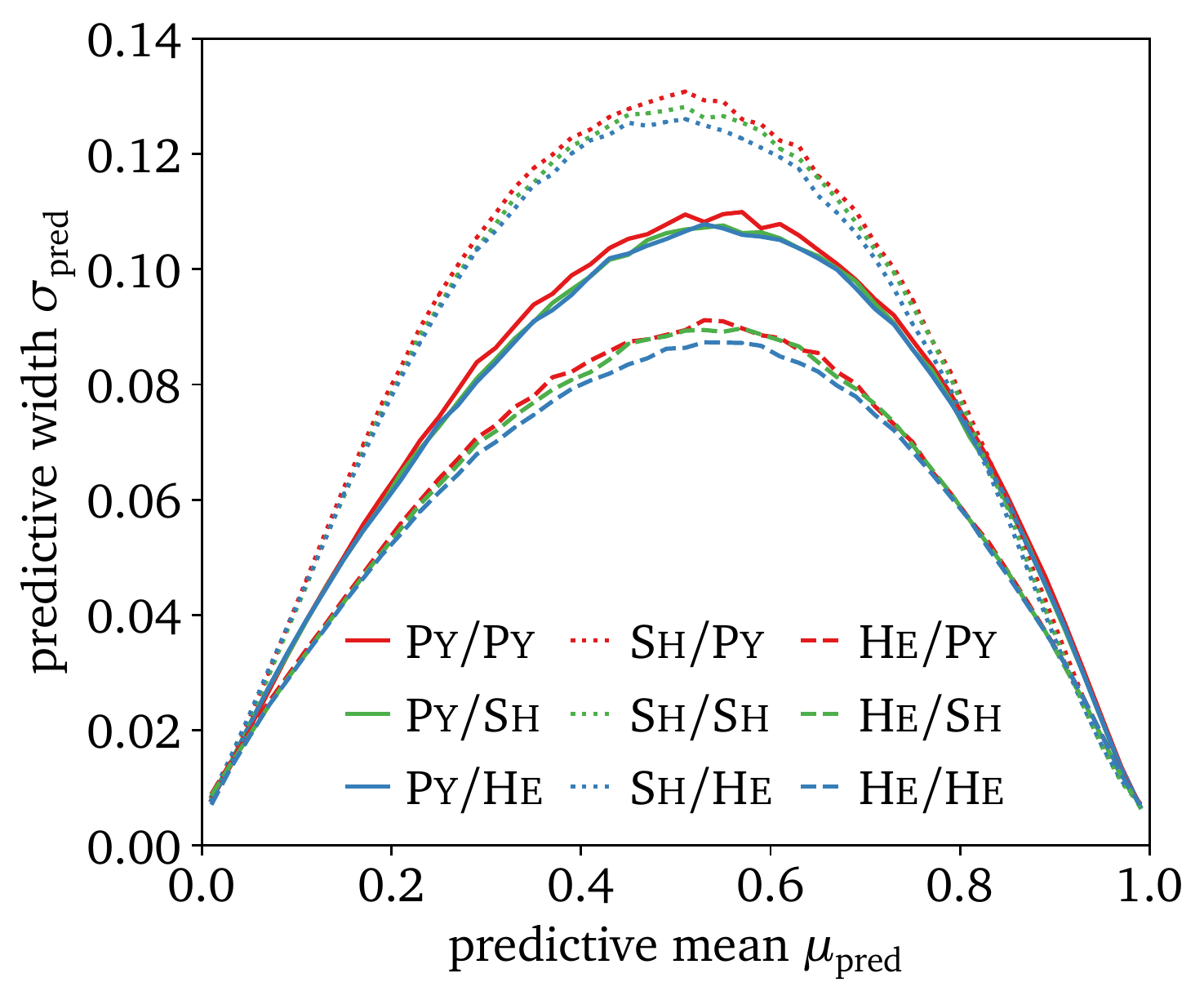}
    \caption{Correlation between the predictive mean and average uncertainty from the BPN-Lite tagger for different combinations of training and testing data.}
    \label{fig:widths}
\end{figure}
%-----------------------------------------------------------

Finally, we can see what the predictive uncertainties tell us in
addition to this information from the network performance. For a given
tagger the predictive mean $\mup$ and the predictive standard
deviation $\sip$ are strongly correlated through
\cref{eq:mu_sigma}, but this argument does not hold for different
training datasets. In \cref{fig:widths} we show the predictive
uncertainties the BPN-Lite tagger extracts when training and testing
on all possible combinations of \pythia, \herwig, and \sherpa. In the
ranges $\mup\sim\range{0.1}{0.9}$ the different training samples define
the size of the predictive uncertainty.

The ranking of the three generators providing the training dataset is
independent of the test sample. This confirms that the predictive
uncertainty of the Bayesian network reflects almost entirely
limitations in the training data. While $\mup$ and $\sip$ are
correlated for a given training dataset, the $\sip$ values in a given
range of $\mup$ are not correlated with the respective $\mup$ values
for different generators. For instance, the poorly performing \herwig
training might not exploit features optimally, but it is less affected
for instance by the stochasticity of the training data. We also see
that any kind of training on \pythia and \herwig will provide smaller
uncertainties on the independent \sherpa data than a Bayesian network
trained on \sherpa and tested on \sherpa. We again emphasize that this
kind of behavior should not appear for $\mup$, because consistent
training should provide better performance than inconsistent training,
but it can happen for $\sip$, as it reflects limitations of the
training dataset only.

%%%%%%%%%%%%%%%%%%%%%%%%%%%%%%%%%%%%%%%%%%%%%%%%%%%%%%%%%%%%%%%%%%%%%%%%%%%%%%
\section{Resilient interpolated training}
\label{sec:resilient}

Once we have understood what the physics issues and the
ML-implications with the \pythia and \herwig training datasets are, we
can follow the setup from the beginning of \cref{sec:where} and
see how to best deal with two significantly different training
datasets, when the tasks is to identify quarks in a third, independent
dataset (\sherpa). This corresponds to the standard ATLAS
and CMS strategy, which is to train ML-classifiers on Monte Carlo
simulations, understand their behavior, and then apply them to
data. The major drawback of this strategy is a generalization error
whenever simulations do not reproduce data perfectly. Such a
generalization error can introduce a bias, but at the very least it is
leading to non-optimal performance. A re-calibration should remove
biases, but it will not improve poorly trained taggers.
We propose a flexible choice of training data, 
defining an optimal training dataset by evaluating the tagger performance on
an independent calibration dataset.

A related question is how to estimate systematic uncertainties
related to the choice of training data. In general, whenever
uncertainties can be described reliably, it is preferable to include
the corresponding nuisance parameters in the analysis, instead of
removing a model dependence through adversarial
training~\cite{Ghosh:2021roe}. Decorrelating theory uncertainties
induced by different datasets is especially tricky, since it enforces
an insensitive direction in feature space and does not allow us to
claim that a general dependence on different training datasets is
significantly reduced~\cite{Ghosh:2021hrh}.

In the case of \herwig \vs \pythia training for quark--gluon tagging
the situation would be even worse, because the two datasets are
systematically different in a way that is fully correlated with the
features used for tagging. Decorrelating the difference of the two
datasets would effectively remove $\npf$ from the available features
and render the tagger useless. Instead, we need to find a way to best
train the tagger and assign an uncertainty to this choice of training
data.

%-----------------------------------------------------------
\begin{figure}[t]
	\centering
	\begin{subfigure}{0.50\linewidth}
		\centering
		\includegraphics[width=0.94\linewidth]{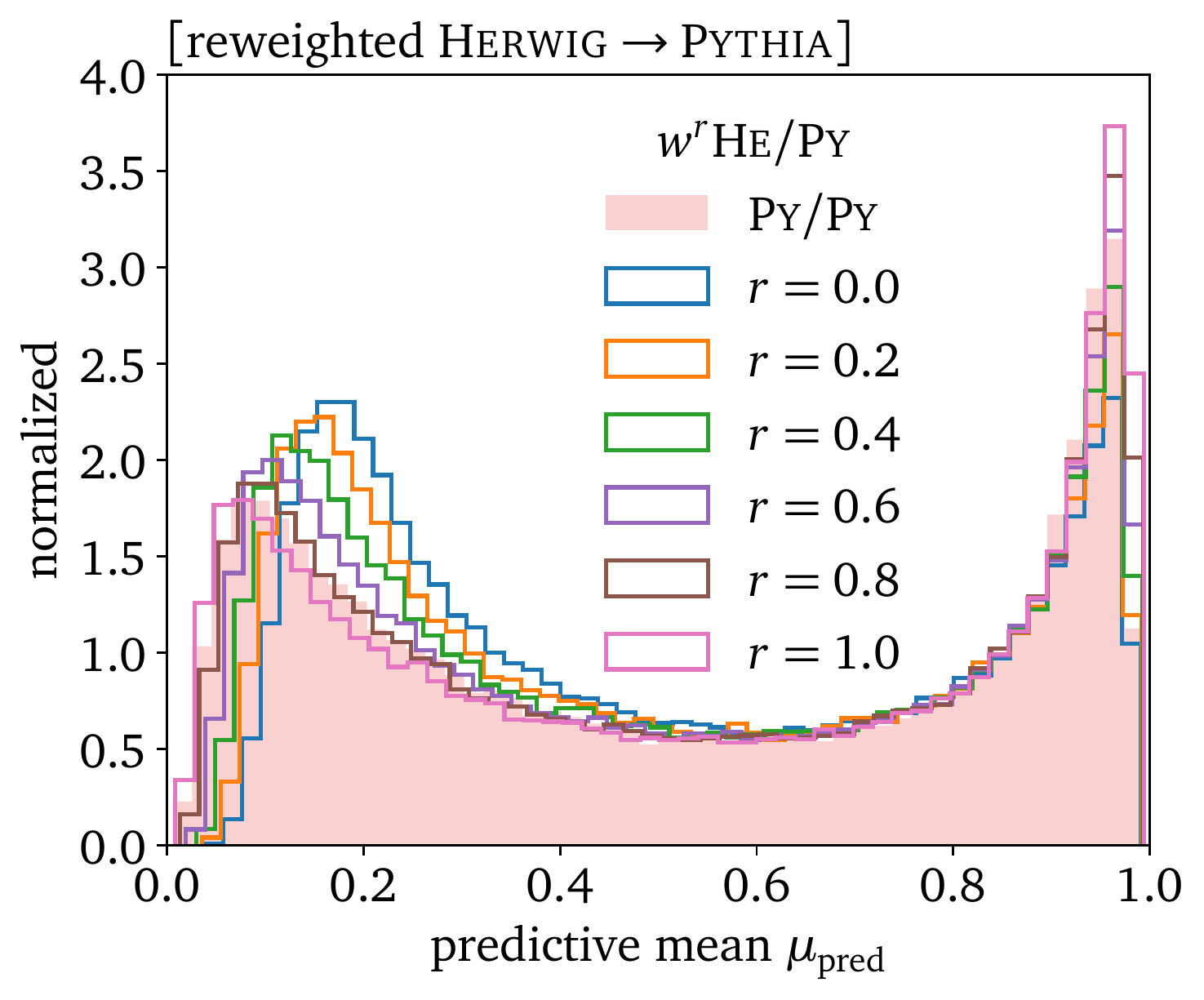}
		%\caption{}\label{fig:}
	\end{subfigure}\hfill
	\begin{subfigure}{0.50\linewidth}
		\centering
		\includegraphics[width=0.94\linewidth]{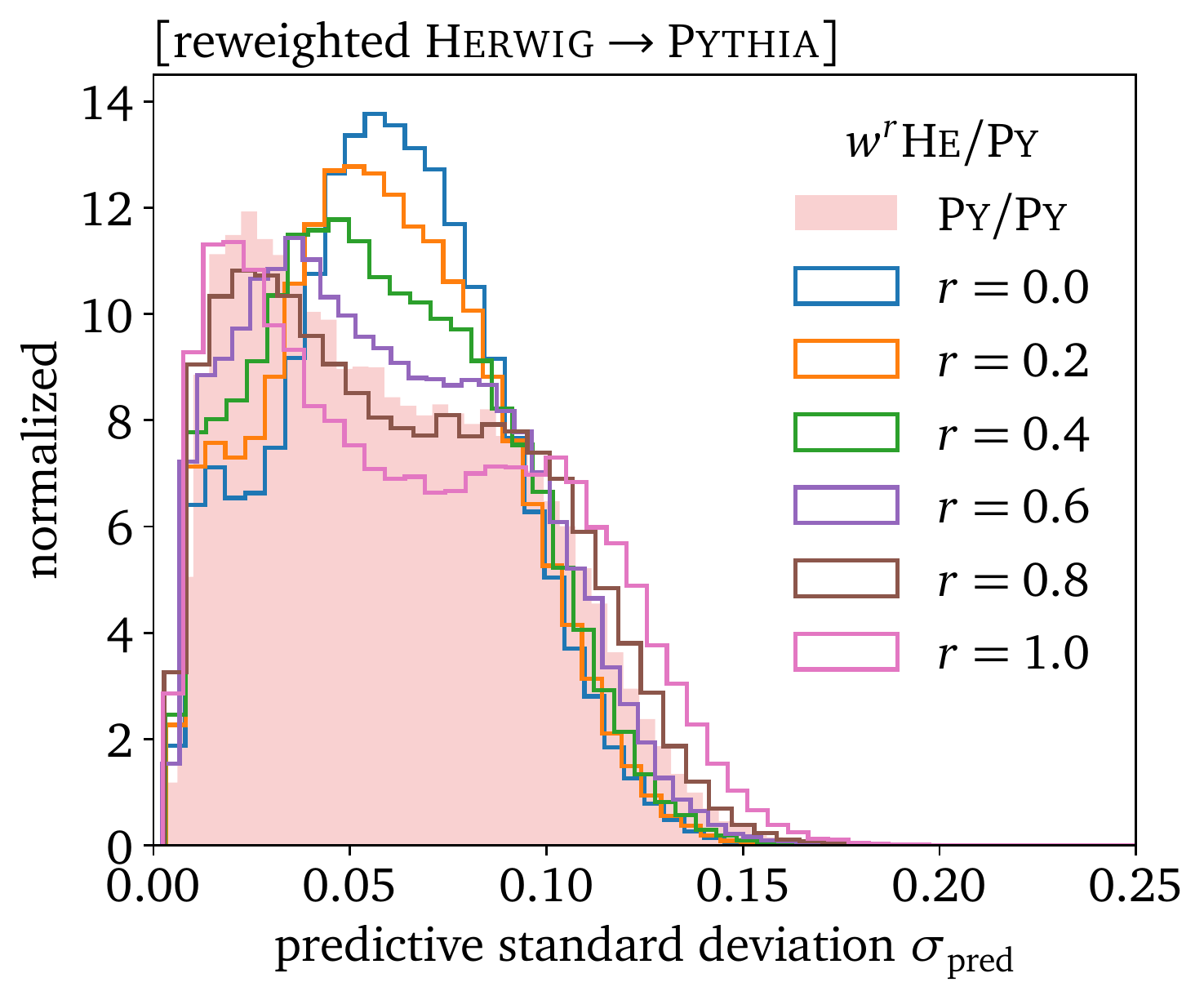}
		%\caption{}\label{fig:}
	\end{subfigure}
    \caption{Bayesian ParticleNet-Lite, trained on reweighted $\herwig\to\pythia$ jets and tested on \pythia jets. The curves should be compared to those in \cref{fig:bpn-mean-std}.}
    \label{fig:interpolate}
\end{figure}
%-----------------------------------------------------------

%%%%%%%%%%%%%%%%%%%%%%%%%%%%%%%%%%%%%%%%%%%%%%%%%%%%%%%%%%%%%%%%%%%%%%%%%%%%%%
\subsubsection*{Interpolated training samples}

To add some resilience to the otherwise extreme choice of training
either on \herwig or on \pythia, we would like to use a combination of
the two datasets for a stable training, benchmarked on the independent
\sherpa data. There are, at least, two ways to interpolate between
the two training datasets. First, we simply train the network on
mixtures of quarks from \pythia and \herwig \vs mixtures of gluons
from \pythia and \herwig in the same proportions,
\begin{equation}
	\text{\herwig training}
	\quad\xlongleftrightarrow{\;\;0\leq r\leq 1\;\;}\quad
	\text{\pythia training} \eqperiod
	\label{eq:interpolate}
\end{equation}
The interpolation parameter $r$ for the mixed sample is the fraction of
\pythia jets in the training dataset.

An alternative method to achieve the same interpolated training is
to train a discriminator on \pythia \vs \herwig quark and gluon jets
and to re-weight the \herwig jets to their \pythia counterparts. Since
each jet now comes with a weight, this method is also only defined on
jet samples. This method has the advantage that we can train the
network conditional on the interpolation parameter $r=\range{0}{1}$, to
stabilize the training. In our case, the discriminator between \herwig
and \pythia jets is the same BPN-Lite network used to tag quarks \vs
gluons. We use the same settings as in \cref{tab:bpn-arch} and the
same loss function as in \cref{eq:loss-bpn}. The only difference
it that for the \herwig \vs \pythia case we use generator truth-labels
instead of jet truth-labels. We train the \herwig \vs \pythia
discriminator for quarks and gluons separately.

Using the per-jet reweighting factors from the classification network,
\begin{equation}
	w(x_{i}) = \frac{p_{\py}(x_{i})}{p_{\he}(x_{i})} \eqcomma
\end{equation}
we can train a quark--gluon classifier on $w^{r}$-reweighted \herwig
jets. The weights enter the BPN-Lite loss function of
\cref{eq:loss-bpn} as
\begin{equation}
	\loss_{\mathrm{BPN}} =
	-\frac{1}{M}\sum_{i=1}^{M} w(x_i)^{r}\log p(y_{i}\vert x_{i}, \omega)
	+\frac{1}{2N}\sum_{\text{weights $\omega_{j}$}}
	\bigl(\mu_{j}^{2}+\sigma_{j}^{2}-\log\sigma_{j}^{2}-1\bigr) \eqcomma
\end{equation}
and the reweighting exponent $r$ is used as an additional feature
input, uniformly sampled from $[0,1]$ during training. In
\cref{fig:interpolate} we illustrate how the conditional
reweighting network works on \herwig jets. We show the
distributions of the predictive mean $\mup$ and the predictive
uncertainty $\sip$ for a tagger trained conditionally on the weighted
samples and tested on \pythia jets. In the limit $r\to 1$ the results
approach the consistent \pythia training and testing shown in
\cref{fig:bpn-mean-std}.

%-----------------------------------------------------------
\begin{figure}[t]
	\centering
	\begin{subfigure}{0.50\linewidth}
		\centering
		\includegraphics[width=0.94\linewidth]{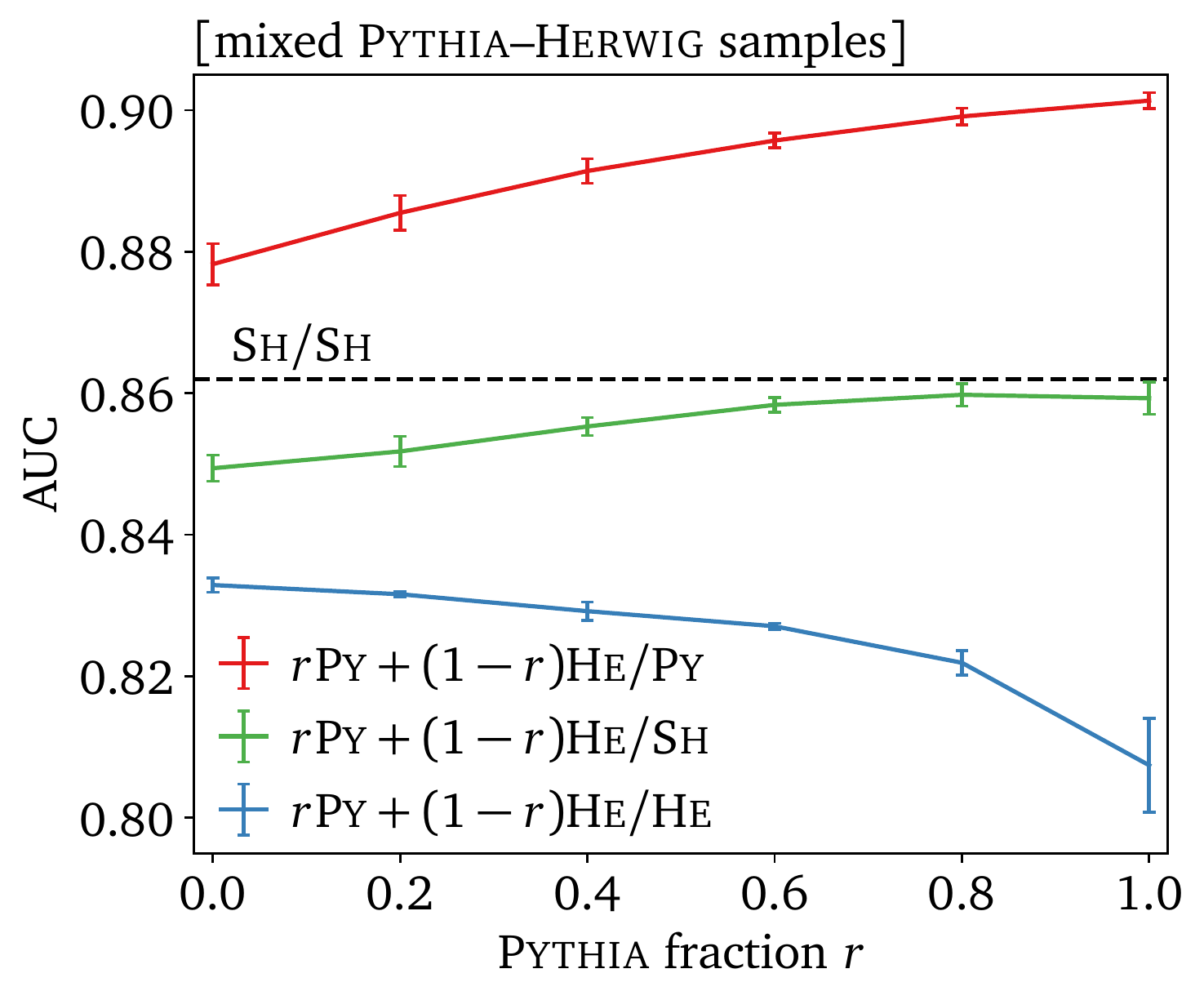}
		%\caption{}\label{fig:}
	\end{subfigure}\hfill
	\begin{subfigure}{0.50\linewidth}
		\centering
		\includegraphics[width=0.94\linewidth]{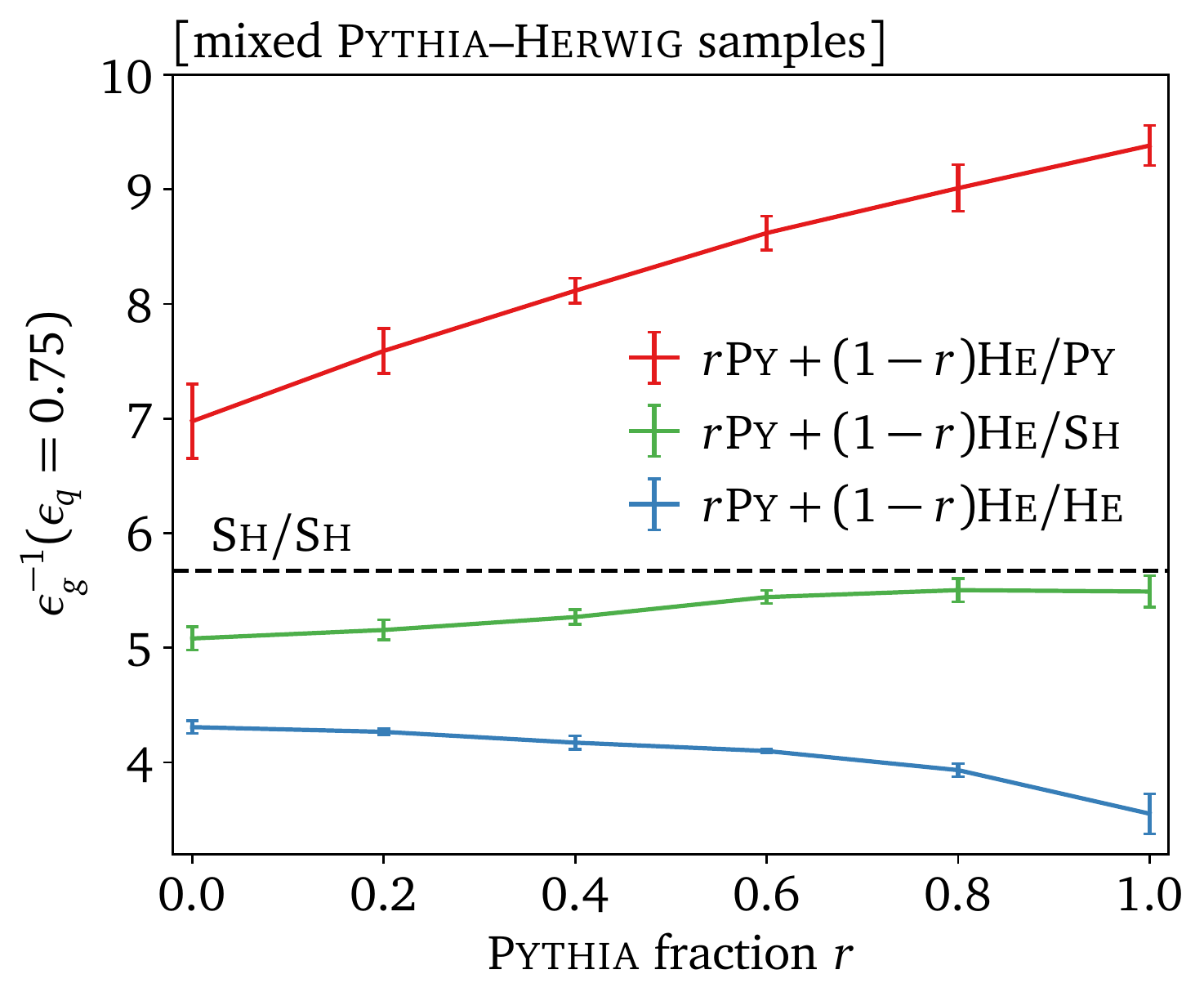}
		%\caption{}\label{fig:}
	\end{subfigure}\newline
	\begin{subfigure}{0.50\linewidth}
		\centering
		\includegraphics[width=0.94\linewidth]{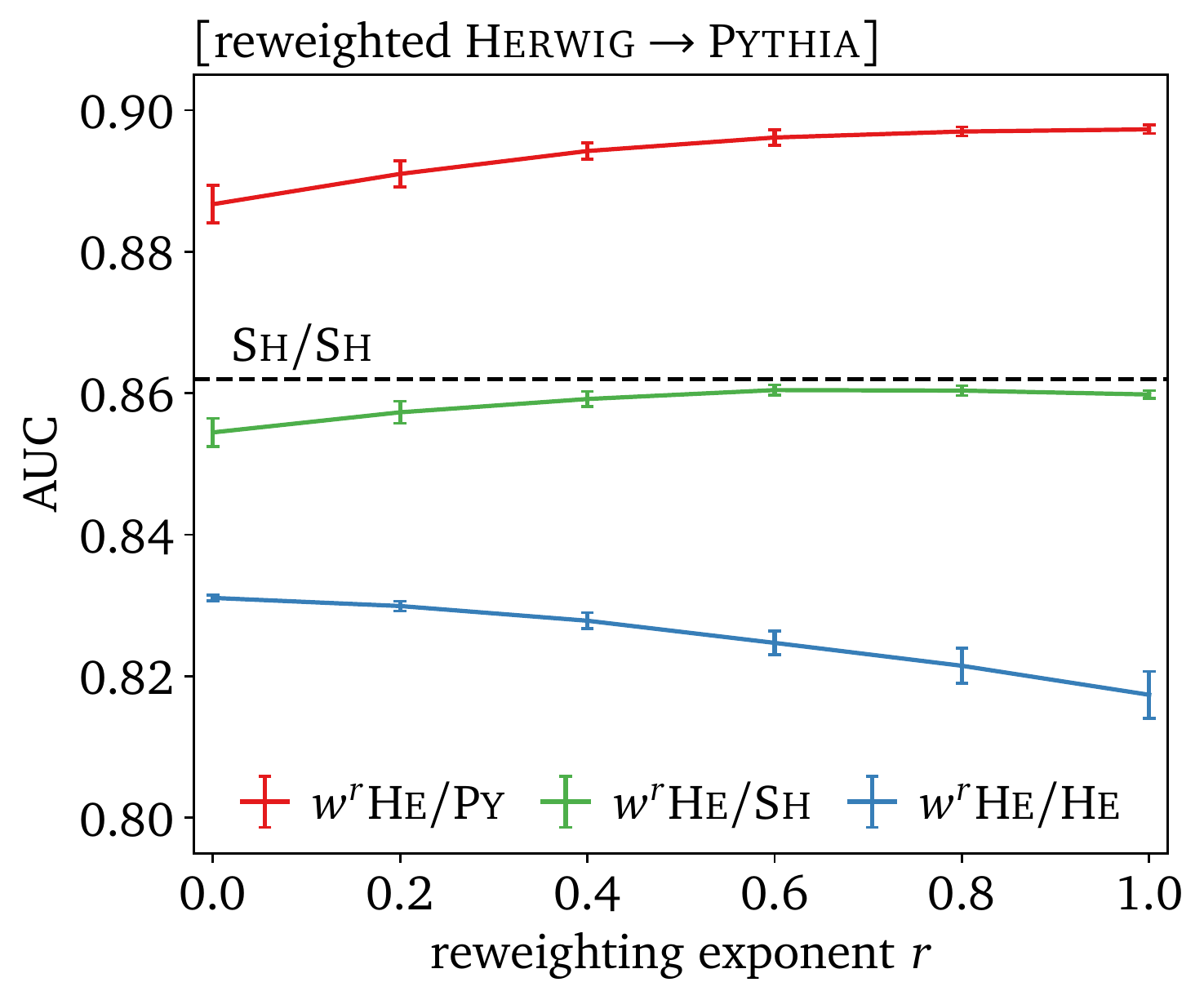}
		%\caption{}\label{fig:}
	\end{subfigure}\hfill
	\begin{subfigure}{0.50\linewidth}
		\centering
		\includegraphics[width=0.94\linewidth]{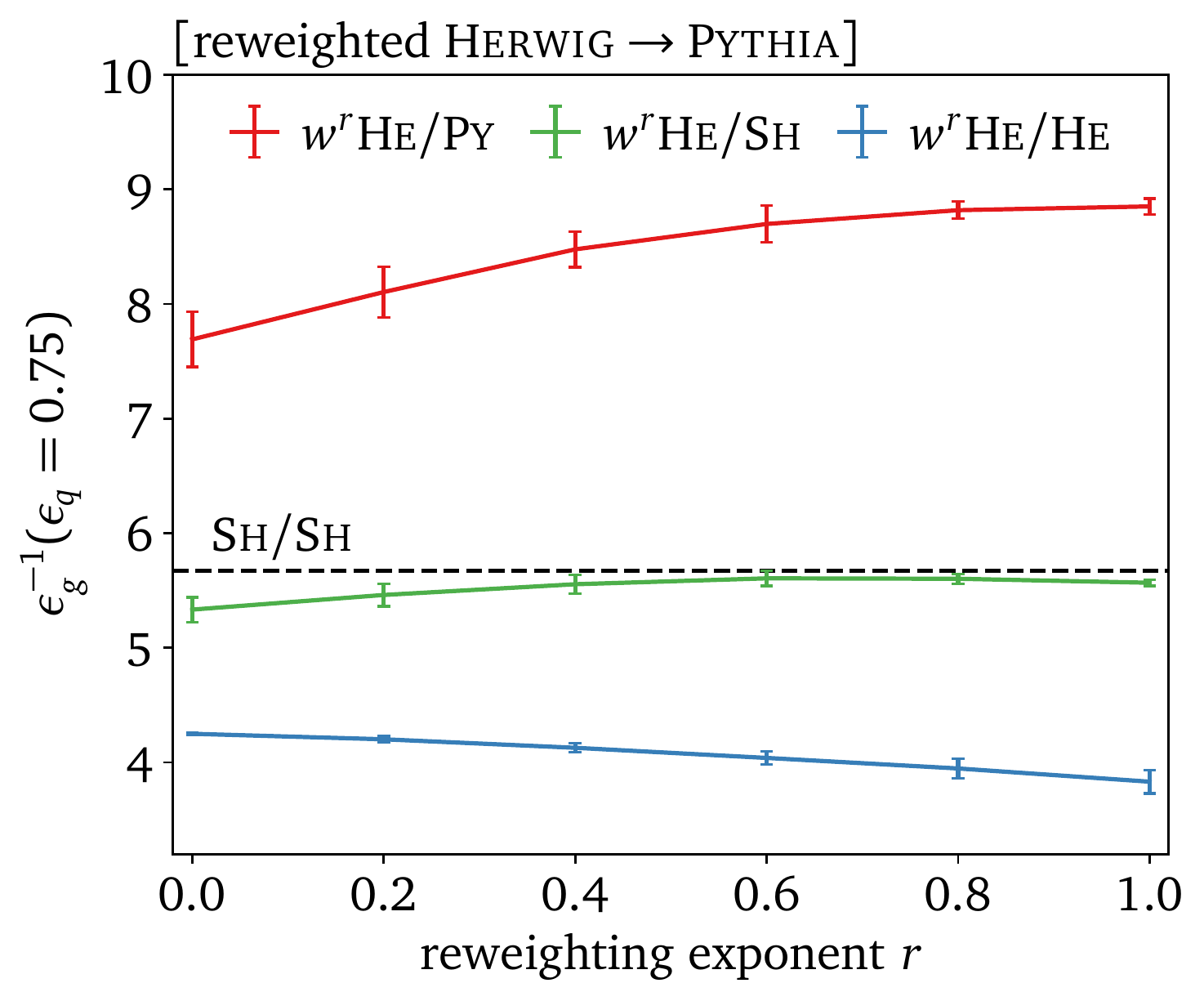}
		%\caption{}\label{fig:}
	\end{subfigure}
    \caption{Performance of the interpolated training on $\herwig \to \pythia$, using mixed samples (\textit{upper}) and conditional reweighting (\textit{lower}). The performance is tested on pure \herwig, \pythia, and independent \sherpa data. The error bars reflect six independent network trainings.}
    \label{fig:mixed_perf}
\end{figure}
%-----------------------------------------------------------

%%%%%%%%%%%%%%%%%%%%%%%%%%%%%%%%%%%%%%%%%%%%%%%%%%%%%%%%%%%%%%%%%%%%%%%%%%%%%%
\subsubsection*{Optimized training data and uncertainties}

The new aspect in this section is the performance of the interpolated
training on the independent \sherpa data. Now, $r$ can be understood
as a hyperparameter of the network training, so we can choose an
optimal value from the independent calibration sample, in our case
\sherpa. The actual tagging performance of the two methods of
interpolated training is shown in \cref{fig:mixed_perf}, with
mixed samples in the upper panels and reweighting in a conditional
network setup in the lower panels. First, the results of the two
methods are completely consistent with each other. They are also
consistent with our previous results; for $r=1$ the performance on all three test datasets approaches the
results from proper \pythia training in \cref{fig:all_auc}. Similarly, in the limit $r=0$ we
reproduce the performance of \herwig-based training. In between, we
observe a continuous and featureless performance drop from \pythia to
\sherpa training. Comparing the two methods, the conditional
reweighting setup is smoother than the mixed sample training.
Finally, after optimizing the interpolated training sample, the
achieves performance comes very close to the consistent \sherpa
training and testing.

%-----------------------------------------------------------
\begin{figure}[t]
	\centering
	\begin{subfigure}{0.50\linewidth}
		\centering
		\includegraphics[width=0.94\linewidth]{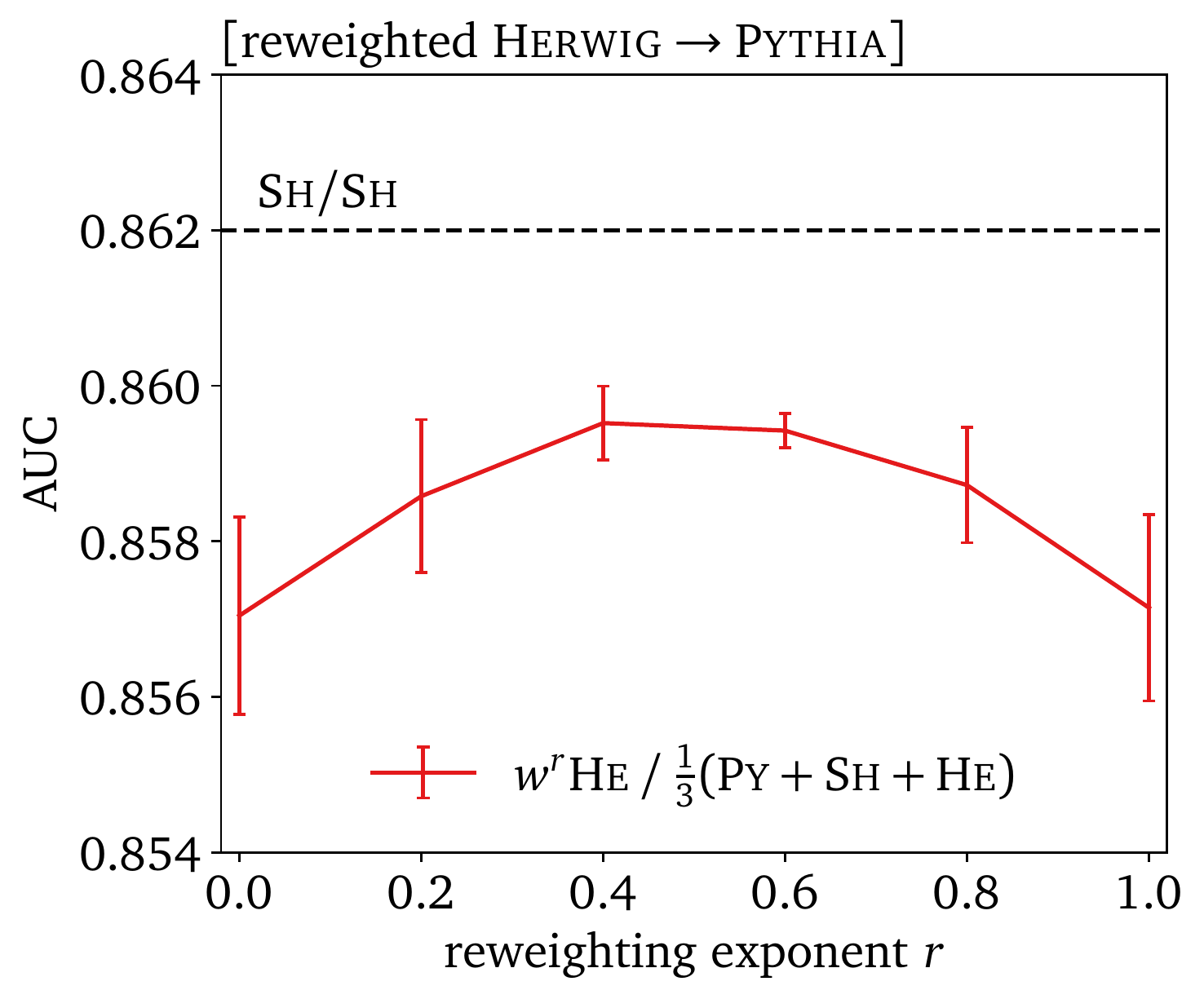}
		%\caption{}\label{fig:}
	\end{subfigure}\hfill
	\begin{subfigure}{0.50\linewidth}
		\centering
		\includegraphics[width=0.94\linewidth]{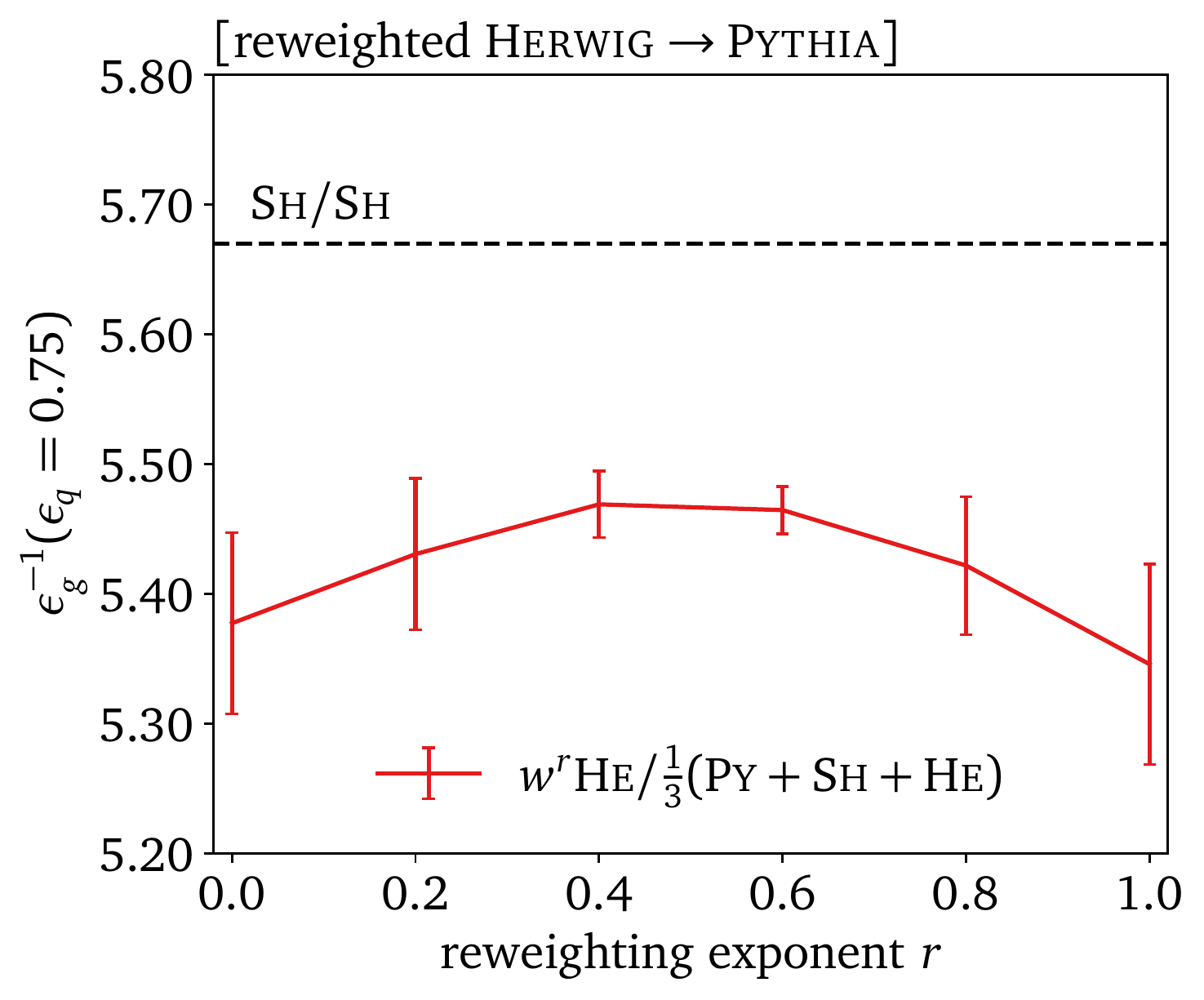}
		%\caption{}\label{fig:}
	\end{subfigure}
    \caption{Performance of the interpolated training on $\herwig\to\pythia$, using conditional reweighting. The performance is tested on equal parts of \herwig, \pythia, and \sherpa jets. The error bars reflect six independent network trainings.}
    \label{fig:mixed_perf2}
\end{figure}
%-----------------------------------------------------------

As a side remark, while testing on \sherpa jets leads us to conclude
that a choice $r\to 1$ provides the optimal tagging performance, we
can also test the interpolated training on combination of \herwig,
\pythia, and \sherpa jets. Because the power of the main tagging
features in the \sherpa dataset tends to lie in between \herwig and
\pythia, shown in \cref{tab:wasserstein-distance}, an interpolated
training with $r\approx 0$ now gives the best tagging performance.

After optimizing the performance on a calibration dataset, we can also 
vary the interpolation parameter $r$ around its
optimal value to estimate the uncertainty from our parameter choice. 
In the lower panels of \cref{fig:mixed_perf} we see that for our 
setup the uncertainty from optimizing in the range $r\approx\range{0.5}{1.0}$
are significantly smaller than the variation from different network trainings.
Strictly speaking, the performance gap even of the best training on the 
combined \pythia and \herwig sample is significant, gauged by the 
uncertainty from the choice of $r$ and from different trainings.
While our example interpolates between two samples, this kind of 
uncertainty estimate can
easily be generalized to many training setups with a conditional
reweighting network.

%-----------------------------------------------------------
\begin{figure}[t]
    \centering
    \includegraphics[width=0.495\linewidth]{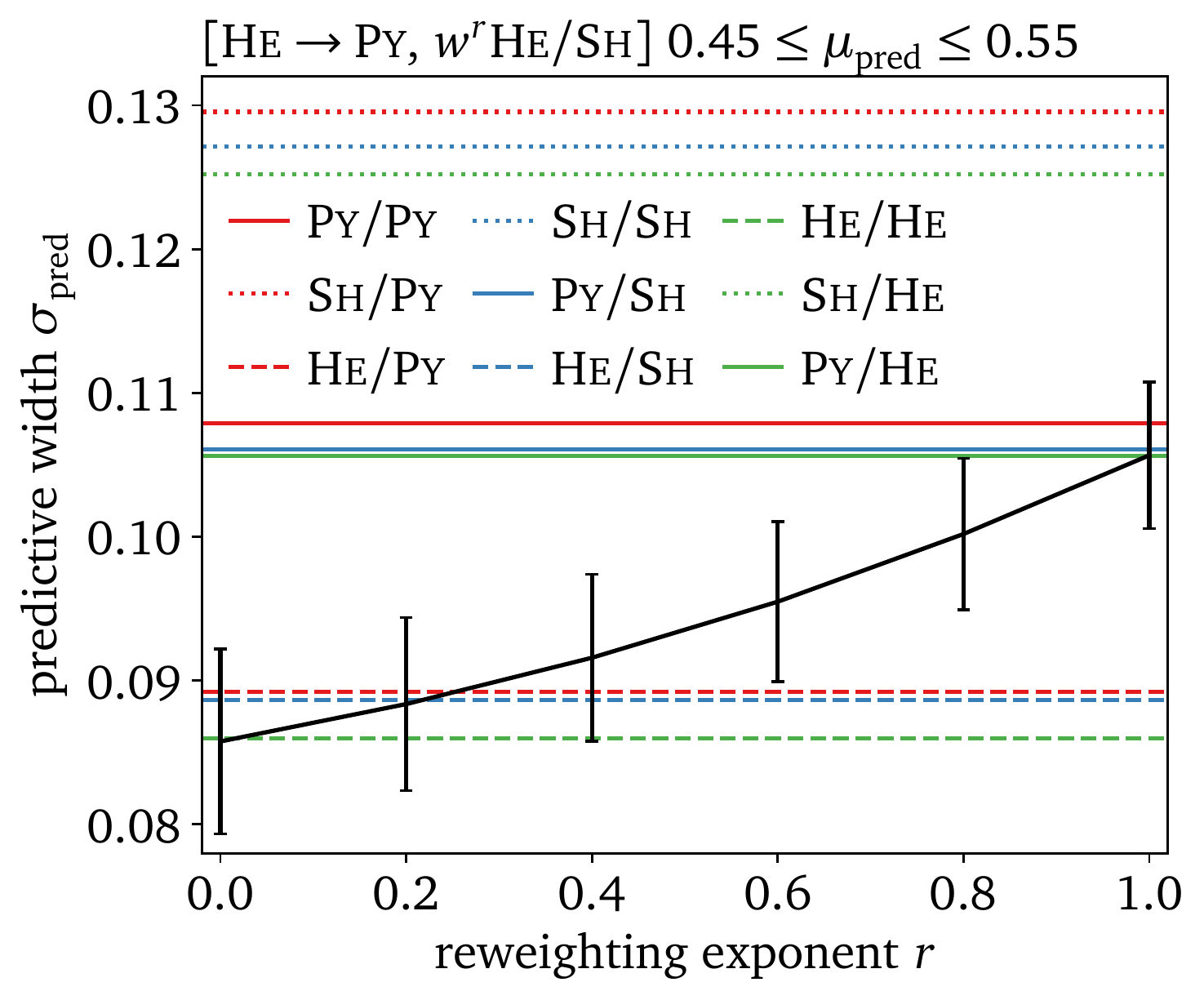}
    \caption{Predictive width for interpolated training on $\herwig\to\pythia$, using conditional reweighting. The error bars indicate the ranges from six independent trainings.}
    \label{fig:mixed_width}
\end{figure}
%-----------------------------------------------------------

%%%%%%%%%%%%%%%%%%%%%%%%%%%%%%%%%%%%%%%%%%%%%%%%%%%%%%%%%%%%%%%%%%%%%%%%%%%%%%
\subsubsection*{Training-related, predictive uncertainties}

We can make use of the uncertainty-aware BPN-Lite tagger to provide
the uncertainties $\sip$ for the interpolated training shown in
\cref{fig:mixed_width}. In analogy to the performance test in
\cref{fig:mixed_perf} we now show $\sip$ as a function of the
interpolation parameter $r$. We know from \cref{fig:widths} that
the predictive uncertainties are given by the training data, and we
can confirm that the interpolated training reproduces the small
\herwig uncertainties for $r=0$ and the slightly larger \pythia
uncertainties for $r=1$. The reweighted and less consistent sample
does not pose a challenge to the training, and the induced
generalization errors are not large enough to affect the results for
the different test datasets. As alluded to before, the interpolated
training on \pythia and \herwig comes with smaller uncertainties than
consistent training on \sherpa, even when tested on \sherpa data. This
can make sense, if the predictive uncertainties just reflect
limitations in the training, for instance noise or stochasticity.

%%%%%%%%%%%%%%%%%%%%%%%%%%%%%%%%%%%%%%%%%%%%%%%%%%%%%%%%%%%%%%%%%%%%%%%%%%%%%%
\subsubsection*{Calibration and uncertainties}

\begin{figure}[t]
	\centering
	\begin{subfigure}{0.50\linewidth}
		\centering
		\includegraphics[width=0.94\linewidth]{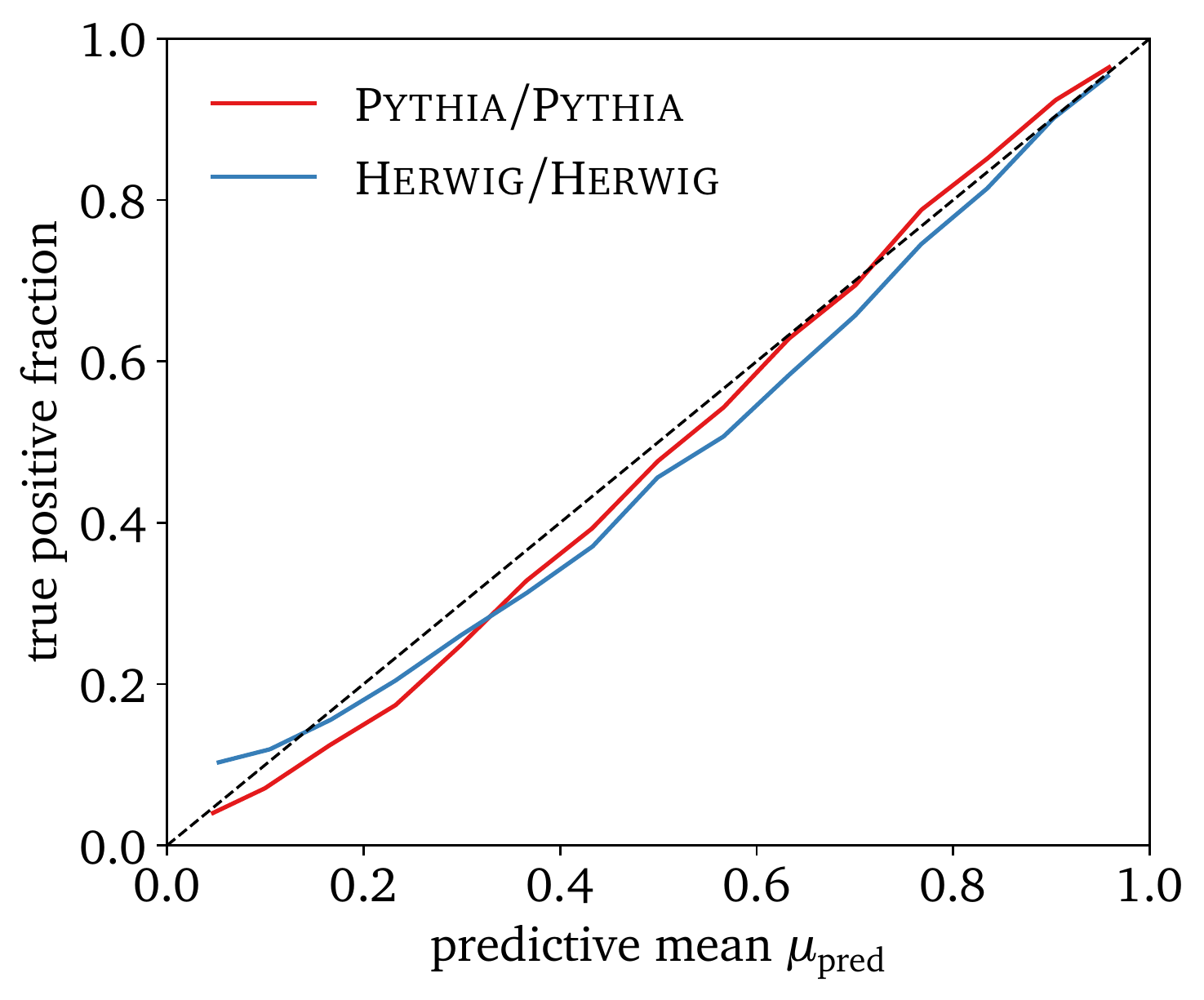}
		%\caption{}\label{fig:}
	\end{subfigure}\hfill
	\begin{subfigure}{0.50\linewidth}
		\centering
		\includegraphics[width=0.94\linewidth]{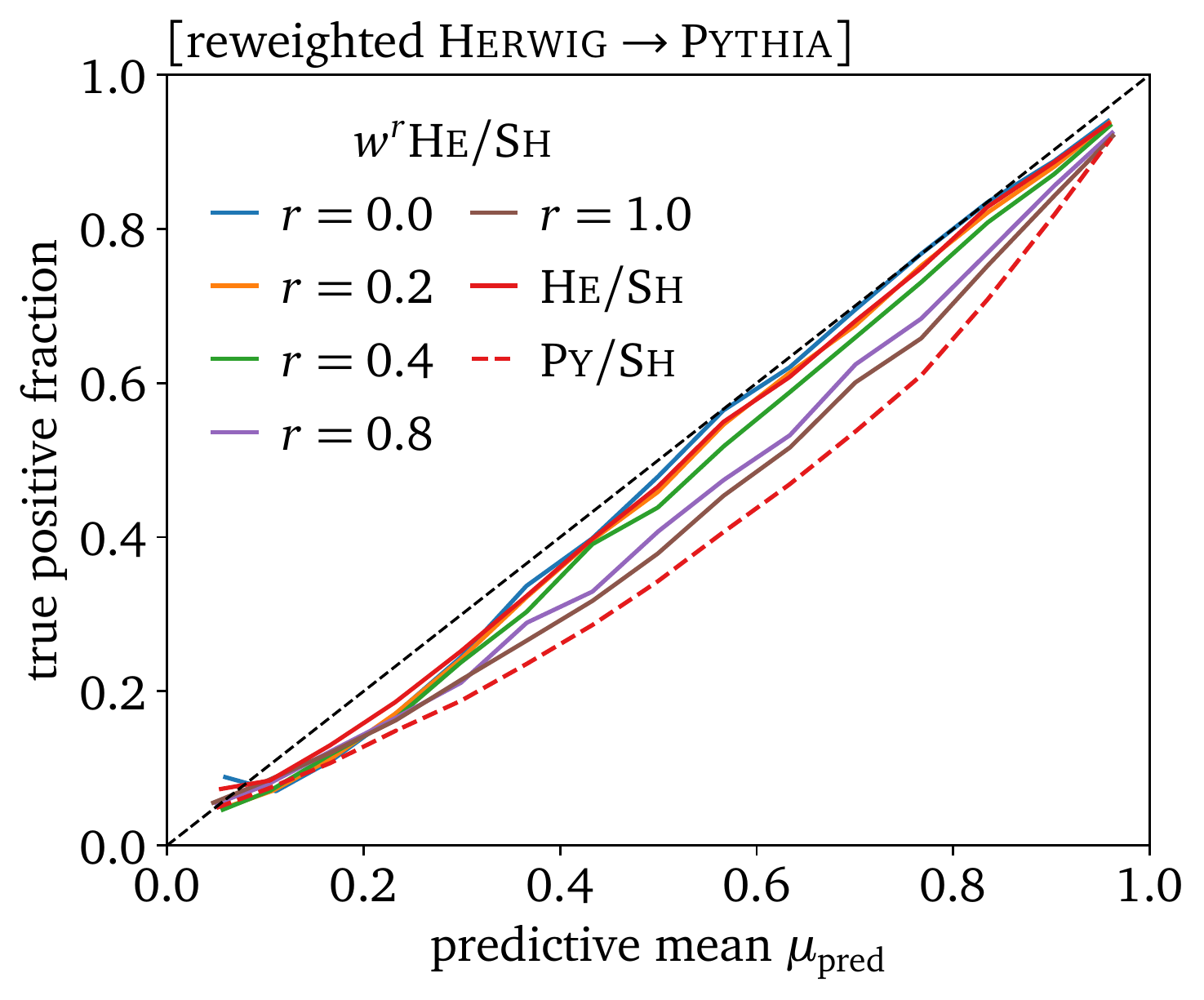}
		%\caption{}\label{fig:}
	\end{subfigure}
    \caption{\textit{Left:} calibration curves for consistent \pythia and \herwig training and testing. \textit{Right:} the same calibration curves using conditional $\herwig\to\pythia$ reweighting on the training data, tested on \sherpa. We also show the benchmark results from training on pure \herwig and \pythia jets, corresponding to $r\in \braces{0,1}$.}
    \label{fig:calibration}
\end{figure}

One measurement where we expect the generalization error to appear is
the calibration of the different taggers. In principle, the Bayesian
PN-Lite tagger should be calibrated, but of course the calibration is
only guaranteed when we train and test on consistent data. Any
deviation from this consistency is expected to lead to a poorer
calibration. In the left panel of \cref{fig:calibration} we first
confirm that the consistent training and testing leads to
well-calibrated taggers over the entire tagging score.

The picture changes when we train the tagger conditionally on the
\herwig--\pythia interpolation and evaluate the calibration on the
independent \sherpa sample. In the right panel of
\cref{fig:calibration} we see that \herwig training leads to a
well-calibrated tagger on the \sherpa dataset, reflecting the fact
that the physics properties behind the two samples are similar. On the
other hand, training on \pythia data leads to a poorly calibrated
tagger on \sherpa data. Here, the fraction of correctly identified
quark jets is lower than the score, which means the tagger is
overconfident. This is consistent with \pythia being the dataset where
it is easiest to separate quarks from gluons.

Because the change in the calibration curve reflects a more dramatic 
$r$-dependence than the network performance in \cref{fig:mixed_perf} 
and the predictive uncertainty in \cref{fig:mixed_width}, it provides 
the best handle on the generalization error which arises when we train
a tagger flexibly on different generated samples and apply it to 
actual (calibration) data.

\bigskip\noindent
To summarize our findings from the interpolated training between
\herwig and \pythia and testing on \sherpa: if we are interested in
the tagging performance only, we need to optimize $r\to 1$,
corresponding to training on pure \pythia jets. When we want to
minimize the Bayesian uncertainties from the training data, training
with $r\to 0$, or on \herwig, will give the smallest predictive
uncertainties. Finally, when we want to maintain the tagger
calibration, we again need to train on $r\to 0$ (\herwig).
Even in ML-applications there is no one size that fits all.

%-----------------------------------------------------------
%-----------------------------------------------------------

%%%%%%%%%%%%%%%%%%%%%%%%%%%%%%%%%%%%%%%%%%%%%%%%%%%%%%%%%%%%%%%%%%%%%%%%%%%%%%
\subsubsection*{Continuous calibration}

If it possible to train a tagger on a continuous interpolation between
different datasets, the same kind interpolation should be possible between
a simulated training dataset and an approximately labeled calibration
dataset. A reliable training dataset should then be transferable into the
actual data continuously, and without major changes in the performance
and the behavior of the ML-tagger. 

Instead of the interpolated training of
\cref{eq:interpolate}, we now look at a triangle, defined by
flexible training on an interpolated \herwig--\pythia dataset and an
additional interpolation between the training data and the independent
\sherpa data,
\begin{equation}
	\text{\herwig/\pythia training}
	\quad\xlongleftrightarrow{\;\;0\leq r\leq 1\;\;}\quad
	\text{\sherpa training} \eqperiod
\end{equation}
In this scheme, we can interpolate several ways, within the training
data, as described before, and from any kind of training data to the
calibration data. The calibration data will not be properly labeled,
so we can either rely on an approximate labeling or apply
classification without labels~\cite{Metodiev:2017vrx}. The only tool
required to implement this complex interpolation program is the 
same ParticleNet classification network that is used for the actual 
tagging task.

\begin{figure}[t]
	\centering
	\begin{subfigure}{0.50\linewidth}
		\centering
		\includegraphics[width=0.98\linewidth]{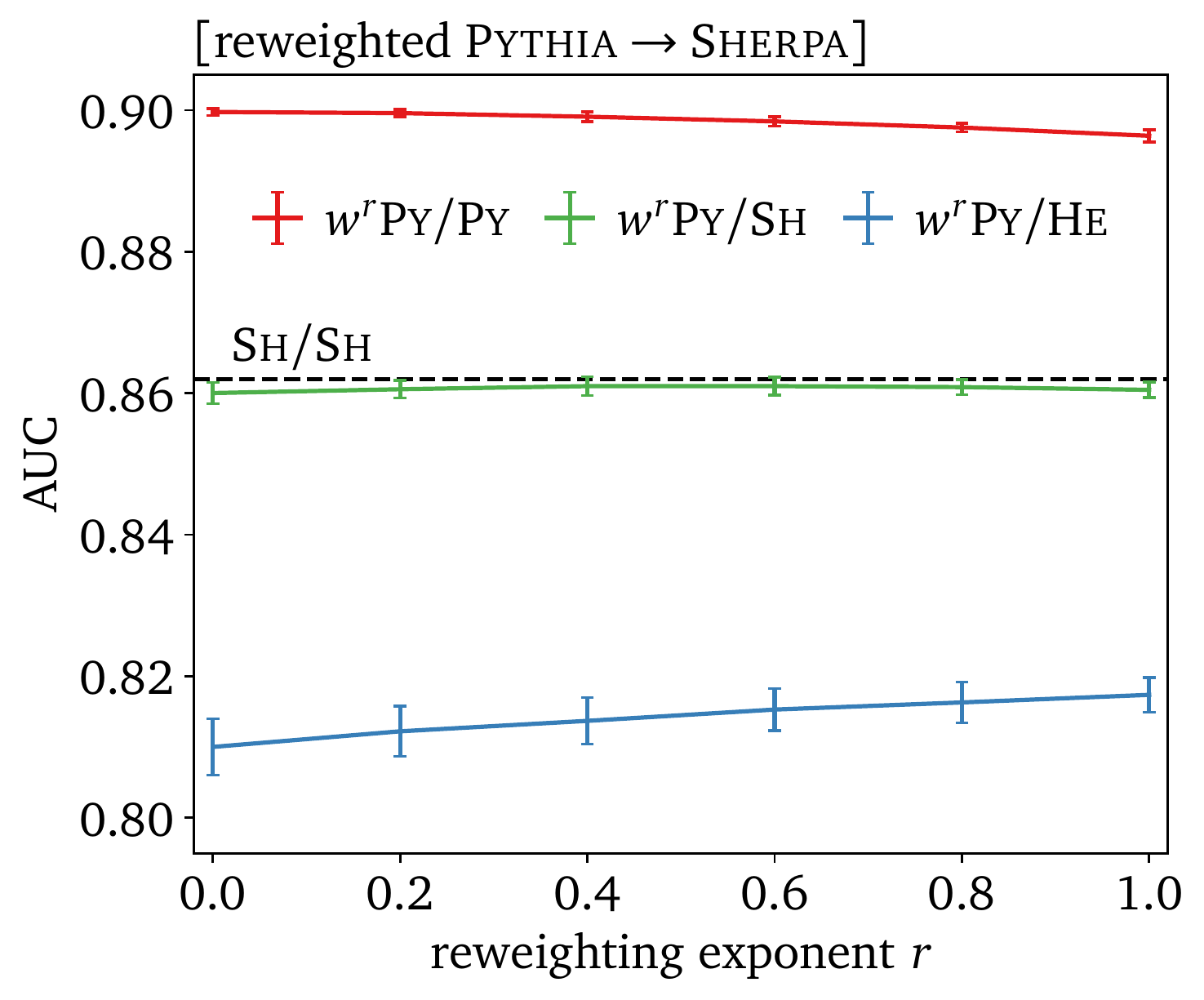}
		%\caption{}\label{fig:}
	\end{subfigure}\hfill
	\begin{subfigure}{0.50\linewidth}
		\centering
		\includegraphics[width=0.98\linewidth]{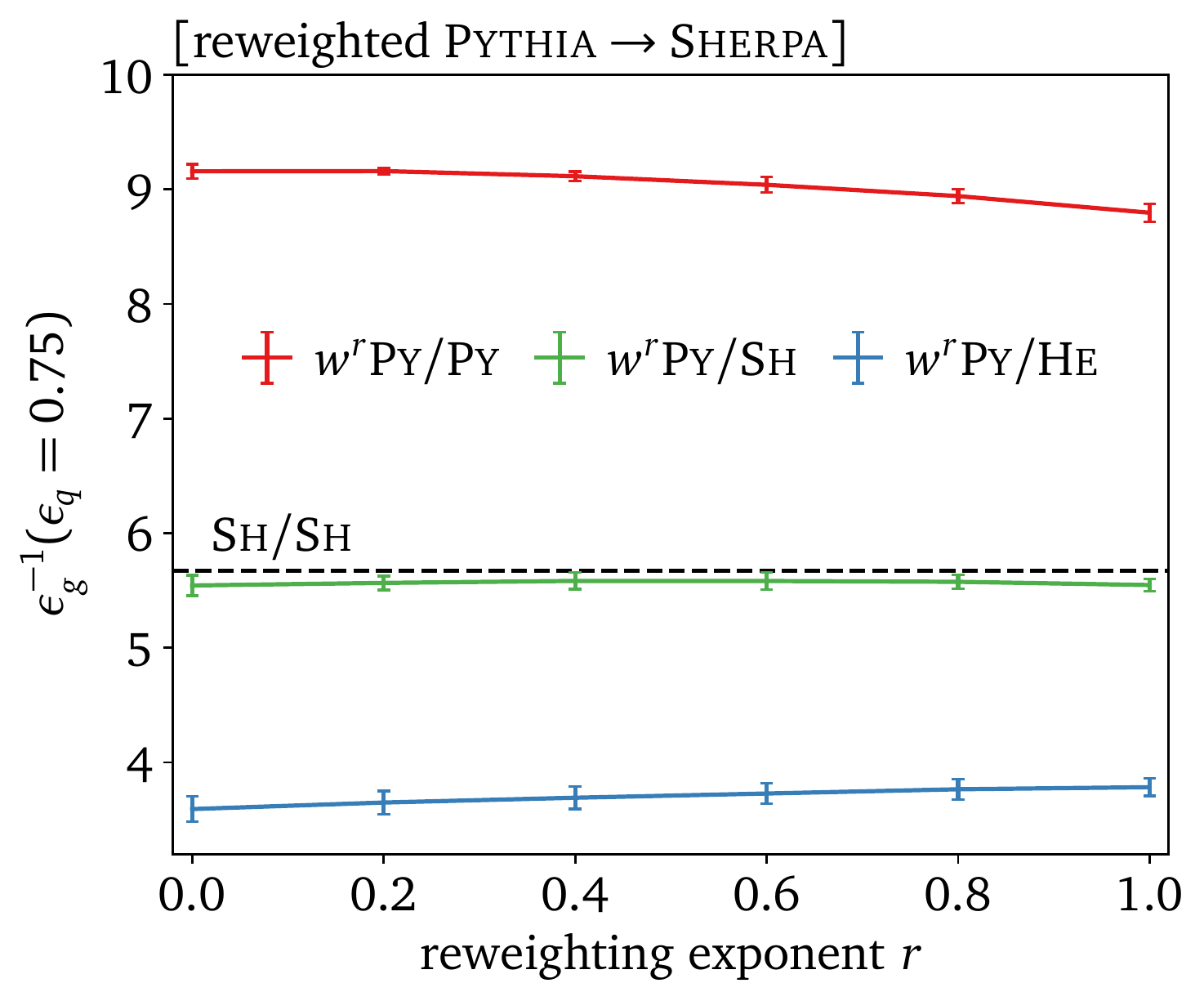}
		%\caption{}\label{fig:}
	\end{subfigure}\newline
	\begin{subfigure}{0.50\linewidth}
		\centering
		\includegraphics[width=0.98\linewidth]{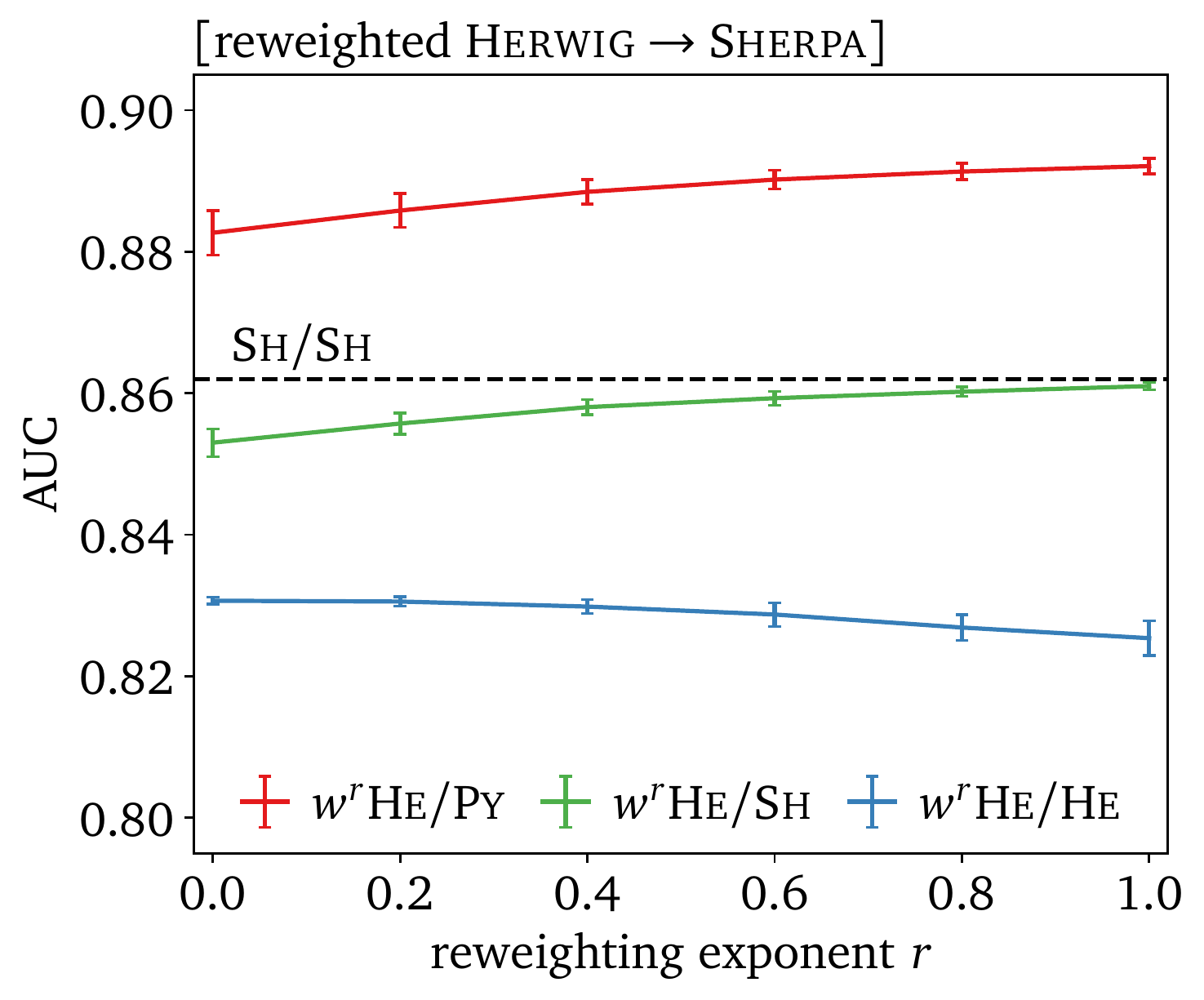}
		%\caption{}\label{fig:}
	\end{subfigure}\hfill
	\begin{subfigure}{0.50\linewidth}
		\centering
		\includegraphics[width=0.98\linewidth]{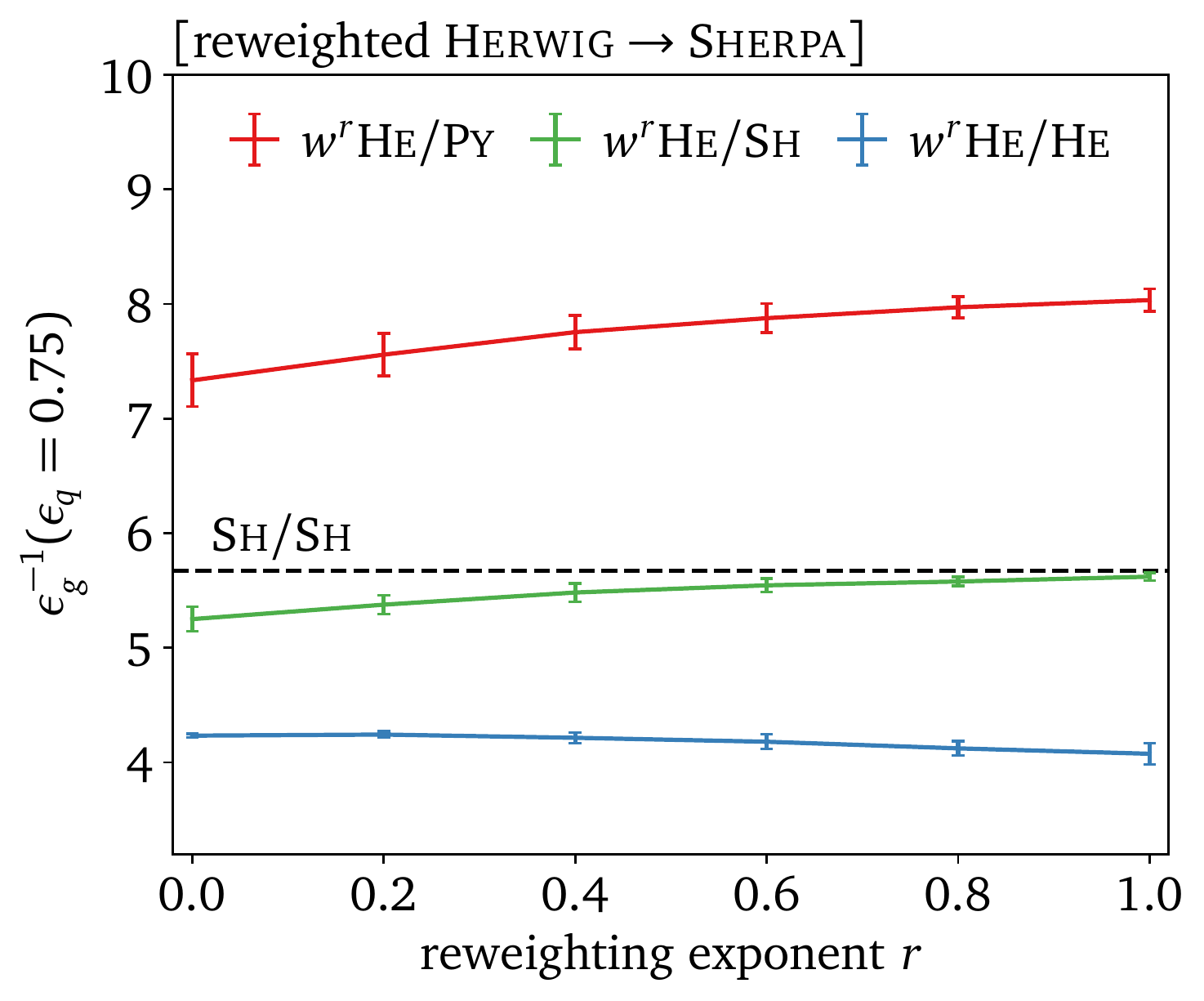}
		%\caption{}\label{fig:}
	\end{subfigure}
  \caption{Performance of the interpolation between training and test
    data, using conditional reweighting $\pythia \to \sherpa$ (\textit{upper}) and
    $\herwig \to \sherpa$ (\textit{lower}). The performance is
    benchmarked on pure \herwig, \pythia, and \sherpa data. The error
    bars reflect six independent trainings.}
    \label{fig:calibration_perf}
\end{figure}

In \cref{fig:calibration_perf} we illustrate the network training
on a continuous interpolation between the simulated data and the
calibration data. The benchmark performance is defined by
the consistent \sherpa training and testing. As expected, there is hardly any change in
performance when we train on \pythia jets, which means that the
calibration of the tagger can focus on the correct calibration. The
situation is different for \herwig training data, where the
performance indicates a generalization gap, but the calibration of the
tagger is stable across the interpolation parameter.
A situation like the one shown in \cref{fig:calibration_perf} 
indicates that training the tagger on simulations (\pythia) and 
applying it to data is, essentially, optimal, with a very small 
uncertainty due to the generalization. 

%%%%%%%%%%%%%%%%%%%%%%%%%%%%%%%%%%%%%%%%%%%%%%%%%%%%%%%%%%%%%%%%%%%%%%%%%%%%%%
\section{Conclusions}
\label{sec:conclusions}

%-----------------------------------------------------------
%-----------------------------------------------------------

Now that it has become clear that ML-jet taggers will provide a 
transformative performance boost to a huge number of
LHC analyses, we should turn our attention to aspects like resilience,
uncertainties, control, or explainability. None of these are
particularly strong points for classic multivariate taggers, so we
again expect ML-taggers to further outperform traditional methods.

As long as we train taggers on simulations and test them on, or apply
them to, an independent dataset, generalization errors will limit their
performance, even if we remove biases through
calibration. These generalization errors contribute to the theory
uncertainty, specifically the dependence of the analysis outcome on
the Monte Carlo simulation.

First, we have shown for quark--gluon tagging based on \herwig and \pythia
training data that improving the resilience through adversarial
training is bound to fail, because the number of constituents is not
only the leading tagging feature, it is also the main difference
between the two simulations.

Just relying on two discrete datasets 
makes it hard to properly evaluate the corresponding theory
uncertainty. We proposed a conditional training on a continuous
interpolation between two training datasets, where the interpolation
is best implemented using re-weighting through a classification
network. The continuous interpolation parameter allows us to
optimize the tagging performance and to estimate the related
uncertainty. Our method can be generalized to larger numbers of
training datasets and to continuous parameters describing the
training datasets.

A Bayesian version of the ParticleNet(-Lite) subjet tagger allows us to
track the stability of the conditional training and identify
training-related uncertainties or even a breakdown of the interpolated
training. For our application to quark--gluon tagging, trained on 
an interpolation between \herwig and \pythia jets and tested on \sherpa
jets, we find that from a pure performance perspective,
training on \pythia gives the best results. They are very close to training
on \sherpa and indicate a very small generalization gap. In
contrast, if we are interested in small predictive uncertainties from
the Bayesian network, we best train on
\herwig data. Similarly, for in a stable calibration \herwig training 
also outperforms \pythia training, reflecting a common physics
picture between \herwig and \sherpa.
For a test dataset combining the three generators, an
interpolated training dataset right in between \herwig and \pythia 
performs best. This indicates that different objectives require
a flexible approach to simulation-based training.

Finally, we have speculated that our continuous interpolation between
training samples can be generalized to an interpolation between
training and calibration data, turning the actual calibration into a
continuous procedure, where stability issues should be easily
detectable.

%%%%%%%%%%%%%%%%%%%%%%%%%%%%%%%%%%%%%%%%%%%%%%%%%%%%%%%%%%%%%%%%%%%%%%%%%%%%%%
\section*{Acknowledgements}

We would like to thank Michel~Luchmann for inspiring discussions on
Bayesian networks, Huilin~Qu for expert advice on jet tagging,
Theo~Heimel for help with the \sherpa dataset, and Mathias~Trabs for 
very helpful advice. A coordinated publication by Markus~Klute and his group,
reflecting inspiring discussions, will appear early in 2023.
Furthermore, we would like to thank Fabrizio~Klassen for his collaboration during an early
phase of this project. AB and TP would like to thank the
Baden-W\"{u}rttemberg-Stiftung for funding through the program
\textsl{Internationale Spitzenforschung}, project
\textsl{Uncertainties -- Teaching AI its Limits}
(BWST\_IF2020-010). BMD acknowledges funding from the Alexander von
Humboldt Foundation. This research is supported by the Deutsche
Forschungsgemeinschaft (DFG, German Research Foundation) under grant
396021762 -- TRR~257: \textsl{Particle Physics Phenomenology after the
  Higgs Discovery} and through Germany's Excellence Strategy
EXC~2181/1 -- 390900948 (the \textsl{Heidelberg STRUCTURES Excellence Cluster}).

\bibliographystyle{SciPost-bibstyle}
\bibliography{literature}
\end{document}

%% file: incl_settings.tex
\usepackage[utf8]{inputenc} % input encodings (allow UTF-8 input)
\usepackage[T1]{fontenc} 	% selecting font encodings (use 8-bit T1 fonts)
\usepackage[english]{babel} % multilingual support (English language/hyphenation)

\usepackage[bitstream-charter]{mathdesign}
\usepackage{geometry} 		% flexible and complete interface to document dimensions
\usepackage{amsmath} 		% math package (American Mathematical Society)
\usepackage{mathtools} 		% math package (fixes various deficiencies of amsmath)
\usepackage{float} 			% floating objects such as figures and tables
\usepackage{graphicx} 		% enhanced support for graphics
\usepackage{tabularx} 		% tabulars with adjustable-width columns
\usepackage{booktabs} 		% professional-quality tables
\usepackage{color, xcolor} 	% foreground and background colour management
\usepackage{pdfpages} 		% inclusion of external multi-page PDF documents
\usepackage{extarrows} 		% extra arrows beyond those provided in amsmath
\usepackage{multirow} 		% create tabular cells spanning multiple rows
\usepackage{multicol} 		% intermix single and multiple columns
\usepackage{caption} 		% customising captions in floating environments
\usepackage{subcaption} 	% support for sub-captions (captions for subfigures)
\usepackage{enumitem} 		% control layout of itemize, enumerate, description
\usepackage{xspace} 		% define commands that appear not to eat spaces
\usepackage{stackrel} 		% enhancement to the \stackrel command
\usepackage{tikz} 			% create PostScript and PDF graphics
\usetikzlibrary{calc}
\usepackage{braket} 		% Dirac bra-ket notation
\usepackage{bm} 			% access bold symbols in maths mode
\usepackage{tensor} 		% typeset tensors (tensor-style super- and subscripts)
\usepackage{slashed} 		% slash through characters (Feynman slash notation)
\usepackage{siunitx} % SI units package (typesetting values with units)
\usepackage{lastpage} 		% reference last page
\usepackage{cite} 			% improved citation handling
\usepackage[normalem]{ulem} % package for underlining
\usepackage{fontawesome} 	% access to web-related icons
\usepackage{tocloft} 		% control over the typography of the TOC, LOF, and LOT
\usepackage{titlesec} 		% interface to sectioning commands (various title styles)
\usepackage{doi} 			% create correct hyperlinks for DOI numbers
\usepackage[most]{tcolorbox} 					% coloured and framed text boxes
\usepackage[capitalize]{cleveref} 	% intelligent cross-referencing
\usepackage[nottoc, notlot, notlof]{tocbibind} 	% add references/index/contents to TOC
\usepackage[ruled, vlined]{algorithm2e} 		% floating algorithm environment
\usepackage{makecell}
\usepackage{pifont}
\usepackage{stackrel}

\hypersetup{
	pdftitle={Performance versus Resilience in Modern Quark-Gluon Tagging},
	pdfauthor={Anja Butter, Barry M. Dillon, Tilman Plehn and Lorenz Vogel},
    linktoc=section,
	breaklinks=true,			% splits links across lines
	colorlinks=true, 			% false: boxed links, true: colored links
	linkcolor={red!50!black}, 	% color of internal links (sections, pages, etc.)
	citecolor={blue!50!black}, 	% color of citation links (links to bibliography)
	urlcolor={blue!80!black} 	% color of URL links (external links)
} 
\DeclareSymbolFont{usualmathcal}{OMS}{cmsy}{m}{n}
\DeclareSymbolFontAlphabet{\mathcal}{usualmathcal}

\SetArgSty{textnormal}
\SetKwComment{Comment}{{\small\#}~}{}
\SetCommentSty{mycommfont}

\setitemize{itemsep=2pt, parsep=0pt} 				% adjust itemize environment
\setenumerate{itemsep=2pt, parsep=0pt} 				% adjust enumerate environment
\crefname{section}{Sec.}{Secs.}
\crefname{equation}{Eq.}{Eqs.}
\crefname{figure}{Fig.}{Figs.}
\crefname{table}{Tab.}{Tabs.}
\Crefname{section}{Section}{Sections}
\Crefname{equation}{Equation}{Equations}
\Crefname{figure}{Figure}{Figures}
\Crefname{table}{Table}{Tables}

%% file: incl_shortcuts.tex
\newcommand{\vs}{vs.\@\xspace} 			% versus (vs.)
\newcommand{\eqcomma}{\;\text{,}} 		% equation comma
\newcommand{\eqperiod}{\;\text{.}} 		% equation period

\newcommand{\braces}[1]{\lbrace#1\rbrace}
\newcommand{\bigbraces}[1]{\bigl\lbrace#1\bigr\rbrace}

\newcommand{\mwith}{\text{with}}
\newcommand{\mand}{\text{and}}

\newcommand{\qqqquad}{\qquad\qquad}

\newcommand{\abs}[1]{\lvert#1\rvert} 		% absolute value (single vertical lines)
\newcommand{\crange}[2]{#1,\ldots,#2}		% range with comma
\newcommand{\range}[2]{#1\:\!\ldots\:\!#2}	% range without comma

\newcommand{\pt}{p_{\mathrm{T}}} 	% transverse momentum
\newcommand{\pti}{p_{\mathrm{T},i}} % transverse momentum (with index i)
\newcommand{\ptj}{p_{\mathrm{T},j}} % transverse momentum (with index i)
\newcommand{\ptjet}{p_{\mathrm{T},\,\mathrm{jet}}}
\newcommand{\etajet}{\eta_{\,\mathrm{jet}}}

\newcommand{\kt}{k_{\mathrm{T}}}
\newcommand{\loss}{\mathcal{L}} 	% loss value

\newcommand{\nc}{n_{C}} % number of constituents
\newcommand{\kl}{\operatorname{KL}}

\newcommand{\npf}{n_{\mathrm{PF}}}
\newcommand{\wpf}{w_{\mathrm{PF}}}
\newcommand{\ptd}{p_{\mathrm{T}}D}

\newcommand{\pid}{\mathrm{PID}}

\newcommand{\auc}{\mathrm{AUC}} % area under the ROC curve
\newcommand{\imtafeqg}[1]{\epsilon_{g}^{-1}(\epsilon_{q}\!=\!#1)}

\newcommand{\pytorch}{\textsc{PyTorch}\xspace}
\newcommand{\tensorflow}{\textsc{TensorFlow}\xspace}

\newcommand{\adamw}{\textsc{AdamW}\xspace}

\newcommand{\fastjet}{\textsc{FastJet}\xspace}
\newcommand{\pythia}{\textsc{Pythia}\xspace}
\newcommand{\herwig}{\textsc{Herwig}\xspace}

\newcommand{\python}{\textsc{Python}\xspace}

\newcommand{\sherpa}{\textsc{Sherpa}\xspace}

\newcommand{\py}{\textsc{Py}}
\newcommand{\he}{\textsc{He}}

\newcommand{\mup}{\mu_{\mathrm{pred}}}
\newcommand{\sip}{\sigma_{\mathrm{pred}}}

\newcommand{\arXiv}[2][]{%
	\ifthenelse{\equal{#1}{}}%
	{\href{http://arxiv.org/abs/#2}{arXiv:#2}}%
	{\href{http://arxiv.org/abs/#2}{arXiv:#2~[#1]}}}

\def\slashchar#1{\setbox0=\hbox{$#1$}           % set a box for #1
   \dimen0=\wd0                                 % and get its size
   \setbox1=\hbox{/} \dimen1=\wd1               % get size of /
   \ifdim\dimen0>\dimen1                        % #1 is bigger
      \rlap{\hbox to \dimen0{\hfil/\hfil}}      % so center / in box
      #1                                        % and print #1
   \else                                        % / is bigger
      \rlap{\hbox to \dimen1{\hfil$#1$\hfil}}   % so center #1
      /                                         % and print /
   \fi}

\newcommand{\tikznode}[2]{%
\ifmmode%
\tikz[remember picture,baseline=(#1.base),inner sep=0pt] \node (#1) {$#2$};%
\else
\tikz[remember picture,baseline=(#1.base),inner sep=0pt] \node (#1) {#2};%
\fi}

\def\mathswitchr#1{\relax\ifmmode{\text{#1}}\else$\text{#1}$\xspace\fi}
\def\mathswitch#1{\relax\ifmmode#1\else$#1$\xspace\fi}